%% file: main.tex
\documentclass[]{aastex631} 

\newcommand\lcdm{$\Lambda$CDM}

\usepackage{amsmath} 

\allowdisplaybreaks 

\shorttitle{GGSL and internal distribution of matter in substructure}
\shortauthors{Tokayer et al.}

\graphicspath{{./}{figures/}}

\begin{document}

\title{The galaxy-galaxy strong lensing cross section and the internal distribution of matter in \lcdm\ substructure}

\input{authors}

\input{abstract}


\section{Introduction} \label{sec:intro}

While the cold dark matter (CDM) model has been incredibly successful in supporting our current best-understanding of structure formation on large scales, challenges to this paradigm on small scales have repeatedly surfaced over the years \citep[e.g.,][]{weinbergColdDarkMatter2015, delpopoloSmallScaleProblems2017, bullockSmallScaleChallengesLCDM2017, perivolaropoulosChallengesLCDMUpdate2022}. 
Tensions between theoretical predictions of the model and observations have arisen as the following problems: the cusp-core issue---simulations predict a universal, cuspy inner density slope of dark matter (DM) halo profiles while some observed galaxies appear to have cores \citep{deblokCoreCuspProblem2010}; abundance of satellites---simulations appear to over-predict the number of expected halo substructures hosting dwarf galaxies compared to observations \citep{mooreDarkMatterSubstructure1999, klypinWhereAreMissing1999}; the predicted kinematics of DM sub-halos coupled with their abundance appear to be inconsistent with data \citep{boylan-kolchinTooBigFail2011}; and the recently reported Galaxy-Galaxy Strong Lensing (GGSL) discrepancy wherein the observed lensing properties of cluster sub-halos are in disagreement with CDM predictions \citep{meneghettiExcessSmallscaleGravitational2020,meneghettiPersistentExcessGalaxygalaxy2023}, which is the subject of this study.

Gravitational lensing---the deflection of light due to space-time deformations by matter---is a powerful probe of DM on all scales, since it is insensitive to the dynamical state of the lens \citep{maoEvidenceSubstructureLens1998, vegettiStrongGravitationalLensing2023}. 
Clusters serve as extremely efficient natural telescopes and via their lensing bring into view faint, background galaxy populations that would otherwise remain undetected. 
There are two regimes in which gravitational lensing effects are observed \citep[for a review, see][]{kneibClusterLenses2011}. 
Strong lensing refers to events that produce arcs, multiple images, and other deformations that are easily discernible, while weak lensing can only be detected statistically by measuring small yet systematic distortions in the shapes of many background galaxies. 
In the strong lensing regime, curves can be discerned along which the most extreme distortions occur for sources that lie behind.
These curves are called critical lines, and they enclose the most concentrated regions of mass in the lens plane.
Thus, lensing configurations trace the detailed spatial distribution of the underlying matter. 
Clusters of galaxies are dominated by dark matter and contain member galaxies that also contain individual bound, dark matter sub-halos. 
Within clusters, dark matter is distributed on both large and small scales, both of which are implicated in the observed lensing effects.

GGSL refers to strong lensing events that occur due to the lensing effect of individual cluster member galaxies and their associated DM sub-halos.
These smaller-scale lensing events are increasingly seen in high-resolution images of cluster lenses such as those that constitute the Hubble Frontier Fields Initiative sample \citep[HFF,][]{lotzFrontierFieldsSurvey2017}.
Individual GGSL events can be identified as perturbations to the large-scale cluster critical lines produced by the smoother dark matter components, and can be used to map the complex mass substructures and their properties in cluster lenses \citep{natarajanLensingGalaxyHaloes1997, natarajanEvidenceTidalStripping2002, natarajanSurvivalDarkMatter2009}.
In particular, GGSL measurements can be used to constrain the sub-halo mass function (SHMF) within clusters, offering a stringent test of the \lcdm\ paradigm on small-scales \citep{dalalDirectDetectionCold2002, keetonNEWCHANNELDETECTING2009, vegettiInferenceColdDark2014, hezavehDETECTIONLENSINGSUBSTRUCTURE2016, gilmanWarmDarkMatter2020, ostdiekExtractingSubhaloMass2022, wagner-carenaImagesDarkMatter2023}.

\cite{meneghettiExcessSmallscaleGravitational2020} (henceforth, M20) reported on the lensing properties of substructure---GGSL---in eleven clusters observed with HST, and demonstrated that there appears to be an order of magnitude inconsistency with mass-matched hydrodynamical simulations of clusters performed with the TreePM-SPH code GADGET-3 \citep{springelCosmologicalSimulationCode2005, beckImprovedSPHScheme2016}.
Simulated sub-halos were found to be under-concentrated and hence unable to match the observed GGSL in these clusters. 
In particular, they measured the GGSL probability, defined as the area enclosed by GGSL caustics in the source plane divided by the full field of view (FoV) of the lens plane projected onto the source plane, as function of source redshift.
They computed the GGSL probability arising from cluster sub-halos with masses in the $10^{10}$-- $10^{11}\ M_\odot$ range.
They find that the observed clusters are over-efficient lenses on small scales compared to their simulated counterparts, with their GGSL probability exceeding them by an order of magnitude or more.
Since sub-halo mass functions derived from lensing models agree with those derived from simulations \citep[e.g.,][]{srivastavaThreeHundredMsub2024,natarajanMappingSubstructureHST2017}, the finding of M20 implies that cluster sub-halos are more centrally concentrated than those predicted in CDM simulations. 

\cite{robertsonGalaxyGalaxyStrong2021} repeated the analysis without implementing the same cuts as M20 for the selection of cluster sub-halos in the higher mass resolution \texttt{C-EAGLE} simulations to compute the GGSL probability.
They included group scale sub-halos with masses $\sim 10^{13}\ M_\odot$ in their census, which automatically increases the GGSL probability and hence claimed that there was no discrepancy given the two orders of magnitude higher mass resolution of the \texttt{C-EAGLE} suite compared to the simulations presented in M20.
However, detailed follow-up studies by \cite{ragagninGalaxiesCentralRegions2022} and \cite{meneghettiProbabilityGalaxygalaxyStrong2022} have pointed out this inconsistent selection of sub-halo masses and hence inappropriate comparison to challenge their claim.
Furthermore, \cite{ragagninGalaxiesCentralRegions2022} and \cite{meneghettiProbabilityGalaxygalaxyStrong2022} demonstrate that even with modifications to the simulations, the estimated GGSL probability from the data cannot be recovered from \lcdm\ simulations while also matching observed properties of cluster galaxies such as their stellar mass and luminosity.
The modifications considered include increasing the mass resolution by factors of ten and 25; significant adjustments to feedback schemes and baryonic physics prescriptions, and changes to the cluster orientation. The subsequent detailed studies demonstrate that mass resolution alone cannot account for the GGSL discrepancy, and in fact increased mass resolution coupled with modified feedback prescription also does not alleviate the issue.

Moreover, it is well known that most, if not all, current simulations are unable to adequately reproduce all of the observed baryonic properties of cluster galaxies \citep{ragone-figueroaBCGMassEvolution2018, ragagninGalaxiesCentralRegions2022}.
\cite{srivastavaThreeHundredMsub2024} similarly tried to address the GGSL discrepancy using simulations from The Three Hundred Project \citep{cuiThreeHundredProject2018} and found that while the SHMF aligns with the lensing observations of M20, the $M_\mathrm{sub}-V_\mathrm{circ}$ relation could not be brought into alignment with any of the baryonic schemes they employed, corroborating the results of M20 and again indicating that observed subhalos are more concentrated than their simulated counterparts. The results reported in \cite{robertsonGalaxyGalaxyStrong2021} are attributable to the inconsistent criterion adopted to classify strong lensing effects by individual galaxies in simulated clusters as demonstrated by \cite{meneghettiProbabilityGalaxygalaxyStrong2022}.

In addition to the finding of M20, unexpected lensing measurements have been reported in observations of field galaxies and Milky Way satellites in recent years.
Perturbations to the lensed arc image around the galaxy SDSSJ0946+1006, for instance, imply the presence of a DM sub-halo. 
Fitting a truncated Navarro-Frenk-White profile to the sub-halo, \cite {minorUnexpectedHighConcentration2021} find that the lensing measurements imply an extraordinarily high halo mass concentration ($\sim10^2$--$10^3$) relative to IllustrisTNG simulation analogs.
More recently, \cite{vandokkumMassiveCompactQuiescent2023} report the discovery of an Einstein ring around a distant ($z\approx2$) galaxy found in the JWST COSMOS-Web survey and show that the mass enclosed within the Einstein radius is higher than what can be accounted for using \lcdm-informed models \citep[though the results of that study have been disputed, see][]{mercierCOSMOSWebRingIndepth2023}.
\cite{andradeHaloDensitiesPericenter2023} studied Milky Way dwarf spheroidal galaxies and report an anti-correlation between satellite orbit pericenter distance and sub-halo mass density, which is in tension with the positive correlation expected from \lcdm\ simulations.

While the majority of the CDM tensions listed above appear to have resolutions within the traditional paradigm, and typically involve taking better account of role of baryons in simulations, occasionally they have catalyzed exploration of alternate DM models, such as self-interacting dark matter (SIDM) to account for the cusp-core issue \citep{spergelObservationalEvidenceSelfInteracting2000}.
Similarly, possible solutions to the GGSL discrepancy fall into two main categories: resolutions within CDM that refine simulations and their predictions \citep[e.g.,][]{ragagninGalaxiesCentralRegions2022}, and resolutions that explore alternative DM models and their effects on halo lensing properties \citep[e.g.,][Dutra et al. in prep]{yangSelfinteractingDarkMatter2021, zengCorecollapseEvaporationTidal2022,amruthEinsteinRingsModulated2023, nadlerSelfinteractingDarkMatter2023}.
Alternate DM models must comport with the large-scale strengths and successful explanatory power of \lcdm\ while addressing the small-scale shortcomings. 
In the context of lensing, alternate DM models may predict different density profiles for cluster sub-halos, thus impacting the overall cluster lensing efficiency.
In this study, we work within \lcdm\ to explore how rearranging the detailed internal distribution of mass within cluster sub-halos effects the computed GGSL, to see if such rearrangements resolve the observed discrepancy.

The key issue in the GGSL tension relates to the detailed distribution of mass within sub-halos.
Even with the current highest resolution HFF images, lens models can only provide integral mass constraints (i.e., total mass within an aperture) from data and are agnostic to how the total matter is configured inside individual sub-halos.
Thus, the detailed distribution of matter within sub-halos is not well-constrained from observations, particularly in cluster environments. 
Working within the \lcdm\ cosmological paradigm, we investigate the impact of altering the detailed distribution of DM and baryons in halo substructures on the GGSL probability.
Note that while the abundance of substructures in mass-matched clusters of \lcdm\ simulations are well-aligned with observations \citep{natarajanMappingSubstructureHST2017}, we are still permitted the freedom to rearrange the mass enclosed inside substructures due to these considerations.
GGSL probability is directly related to the area enclosed by small-scale critical lines.
Since critical lines are produced where the projected mass density reaches a critical value (see Sec.~\ref{subsec:ggsl_def}), rearranging the mass into cuspier profiles may decrease the GGSL probability, bringing the observations into agreement with simulations.
However, since GGSL calculations are highly nonlinear, the GGSL effects of different mass profiles can be difficult to predict without computing them directly.

In this study, we first reproduce the results of M20, which shows the GGSL tension, using the publicly available lensing-derived mass profiles for five observed clusters.
We then rearrange the mass in galaxy-scale sub-halos of those clusters into other profiles consistent with the \lcdm\ paradigm to explore the effect on the computed GGSL probability, and to see if such rearrangements can resolve the GGSL tension.
The five clusters to which we apply this methodology are (see Sec.~\ref{sec:samples}): Abell S1063, MACS J0416.1-2403, MACS J1206.2-0847, and Abell 2744 from the M20 sample, as well as PSZ1 G311.65-18.48 from the sample of \cite{meneghettiProbabilityGalaxygalaxyStrong2022}.

The outline of our paper is as follows: we describe the dataset in Sec.~\ref{sec:samples}.
In Sec.~\ref{sec:methodology} we define GGSL probability and describe our methodology for its computation. 
In Sec.~\ref{sec:ggsl_comparison_results}, we compare the GGSL probability of two separate interior mass profiles for sub-halos.
The role of baryons is addressed and explored in Sec.~\ref{sec:baryons}, and we conclude with a discussion in Sec.~\ref{sec:conc}. 
Throughout this paper, we assume $h=0.69$ and $\Omega_{\mathrm{m},0} = 0.3$, consistent with the WMAP9 results \citep{hinshawNINEYEARWILKINSONMICROWAVE2013}.


\section{Cluster sample} \label{sec:samples}

In this section, we summarize the relevant properties of the five lensing clusters studied here. Four are lensing selected clusters that are part of the HFF program, and which were shown in M20 to be over-efficient GGSL lenses, while one of the clusters, PSZ1 G311.65-18.48, is from the ESA Planck catalog of Sunyaev-Zeldovich (SZ) selected clusters \citep{planckcollaborationPlanck2013Results2014}, and was shown in \cite{meneghettiProbabilityGalaxygalaxyStrong2022} to also be an over-efficient GGSL lens \citep[see also][]{meneghettiPersistentExcessGalaxygalaxy2023}, even though it was selected using completely different criteria from the HFF clusters. 

The mass maps obtained from lensing observations that we use in this study are derived from \texttt{LENSTOOL}, a publicly available lens inversion code \citep{kneibHubbleSpaceTelescope1996, julloBayesianApproachStrong2007, julloMultiscaleClusterLens2009}. 
For each cluster, \texttt{LENSTOOL} returns parameter values for the mass distribution that offer the best fit between the shapes, positions and brightness of the observed multiple images with those predicted by the model. 
The optimized result is a projected mass map that is fully described by parameters that partition the mass into the large-scale cluster halos and the smaller-scale galaxy components.
The \texttt{LENSTOOL} models assume dPIE profiles (see Sec.~\ref{subsec:parametric_models} below) and are publicly available for the four HFF clusters.\footnote{\url{https://archive.stsci.edu/prepds/frontier/lensmodels/}}
The details of the reconstructions themselves can be found in \cite{johnsonLensModelsMagnification2014}, 
\cite{mahlerStronglensingAnalysisA27442018}, \cite{bergaminiEnhancedClusterLensing2019}, \cite{bergaminiNewHighprecisionStrong2021}, \cite{pignataroStrongLensingModel2021}, \cite{meneghettiProbabilityGalaxygalaxyStrong2022}, and the supplementary materials of M20 (henceforth, M20sup). 
In order to reduce the number of free parameters when performing fits to dPIE profiles, \texttt{LENSTOOL} employs luminosity scaling relations for certain parameters of the small scale cluster components (see Sec.~\ref{subsec:parametric_models}).
In the paragraphs that follow we describe each of the clusters analyzed in this work.
Table~\ref{tab:cluster_info} summarizes their pertinent properties.
\input{cluster_samples_table} 

Abell S1063 (AS1063), located at $z=0.348$, was first identified by \cite{abellCatalogRichClusters1989} and was observed as part of HFF as well as the Cluster Lensing And Supernova survey with Hubble program \citep[CLASH,][]{postmanCLUSTERLENSINGSUPERNOVA2012}CLASH.
It was further observed as a CLASH-VLT target \citep{rosatiCLASHVLTVIMOSLarge2014}, and is a year three target of the Beyond Ultra-deep Frontier Fields And Legacy Observations survey \citep[BUFFALO,][]{steinhardtBUFFALOHSTSurvey2020}. 
AS1063 stands out for its high X-ray luminosity, which suggests that it is going through a major merging event \citep[][and references therein]{gomezOPTICALXRAYOBSERVATIONS2012}. 
The cluster has also been studied in the radio \citep{rahamanXrayRadioStudy2021} and optical \citep[][and references therein]{beauchesneNewStepForward2023} bands, as well as via weak lensing \citep{gruenWeakLensingAnalysis2013}.
\cite{williamsonSUNYAEVZELDOVICHSELECTEDSAMPLE2011} estimate the SZ derived mass to be $M_{200}=(2.90\pm1.33)\times10^{15}\ M_\odot\ h^{-1}$. 
The mass models obtained consist of 227 potentials, three of which were high mass ($>5\times10^{12}\ M_\odot$) and another two that were also optimized outside of the scaling relations.

MACS J0416.1-2403 (M0416) is located at $z=0.397$ and was discovered by the Massive Cluster Survey (MACS) of ROSAT X-ray selected clusters \citep{ebelingMACSQuestMost2001}. 
It has been imaged by HST as part of HFF and CLASH, was further observed as part of CLASH-VLT, and is a year one BUFFALO target.
In the context of the BUFFALO, \cite{gonzalezSettingSceneBUFFALO2020} measured the surface mass density of M0416 from a weak-lensing analysis and obtained a higher number of sub-halos when compared to a simulated cluster of similar mass.
M0416 is in a complex dynamical state, featuring a bimodal distribution and elongation of the main substructures, suggesting a pre-collisional stage \citep{balestraCLASHVLTDISSECTINGFRONTIER2016}.
\cite{jauzacHubbleFrontierFields2014} use the HFF image data of M0416, which includes 194 lensed images of 68 background galaxies, to constrain the projected mass enclosed within 140 kpc $h^{-1}$ to be $(1.12\pm0.01)\times10^{14}\ M_\odot\ h^{-1}$. The CLASH analysis presented in \cite{umetsuCLASHJOINTANALYSIS2016} estimates $M_\mathrm{vir} = (0.909\pm 0.230)\times10^{15}\ M_\odot\ h^{-1}$ for the cluster.
The mass model reconstruction of M0416 consists of 197 potentials, six of which are large-scale components optimized outside of the scaling relations.

MACS J1206.2-0847 (M1206), located at $z=0.440$, is also included in the MACS survey of X-ray luminous clusters, and was observed as part of CLASH-VLT. 
While X-ray emission suggests an overall relaxed state, much of the intracluster light is not centrally concentrated, suggesting ongoing galaxy-scale interactions \citep{eichnerGALAXYHALOTRUNCATION2013}. 
Among its 82 spectroscopically confirmed multiple images is a prominent 15'' long gravitational arc west of the central galaxy \citep{sandDarkMatterDistribution2004}. 
Assuming a spherical NFW halo, \cite{umetsuCLASHMassDistribution2012} estimate the virial mass of M1206 to be $M_{132} = (1.1\pm0.3)\times10^{15}\ M_\odot\ h^{-1}$. 
\cite{richardAtlasMUSEObservations2021} perform integral field spectroscopy analysis of M1206 multiple images using data from the Multi Unit Spectroscopic Explorer (MUSE) instrument on VLT, and using \texttt{LENSTOOL} models, they measure the mass within the Einstein radius of the cluster (for a source plane at $z_\mathrm{s}=4$) to be $(1.23\pm0.04)\times10^{14}\ M_\odot\ h^{-1}$. 
The mass models obtained consist of 265 potentials, six of which are high mass halos optimized outside the scaling relations, and an additional potential representing the cluster shear field within the FoV.

Abell 2744 (A2744), at $z=0.308$, is another merging cluster, nicknamed ``Pandora's cluster'' by \cite{mertenCreationCosmicStructure2011} due to its exciting complexity. 
For example, it has the largest offset between X-ray and lensing centers among the 38 clusters studied by \cite{shanOffsetDarkMatter2010}. 
The cluster has been extensively studied from radio to X-ray bands, is part of the HFF program, the Ultra-deep NIRSpec and NIRCam ObserVations before the Epoch of Reionization program \citep[UNCOVER,][]{bezansonJWSTUNCOVERTreasury2022} with JWST, and is a year one BUFFALO target. One of the galaxies lensed by A2744 was recently confirmed using combined Chandra and JWST observations to harbor the most massive distant ($z>10$) accreting black hole observed to date \citep{bogdanEvidenceHeavyseedOrigin2023}.
The HFF strong lensing analysis by \cite{jauzacHubbleFrontierFields2015} measured the projected mass within a 140 kpc $h^{-1}$ radius to be $(1.513\pm0.004)\times10^{14}\ M_\odot\ h^{-1}$, and strong and weak lensing measurements by \cite{jauzacExtraordinaryAmountSubstructure2016} constrain the projected mass within a 0.9 Mpc $h^{-1}$ radius to be $(1.6\pm0.1)\times10^{15}\ M_\odot\ h^{-1}$.
The mass models obtained consist of a total of 258 potentials, six of which were large scale potentials, and another six that represent lensing potentials beyond the FoV needed in order to best fit the observed lensing.

PSZ1 G311.65-18.48 (PG311) is located at $z=0.443$ and was first confirmed as a galaxy cluster as part of the ESA Planck catalog of SZ selected clusters \citep{planckcollaborationPlanck2013Results2014}.
We include this cluster in our sample to explore the applicability of our results to clusters that are not selected via strong-lensing, in case that selects preferentially for over-concentrated halos \citep{oguriSubaruWeakLensing2009, grallaSunyaevZelDovichEffect2011}. 
Its most notable feature is a striking tangential arc image of a background star forming galaxy at $z=2.369$ that is imaged at least 11 other times, dubbed the ``Sunburst Arc'' by \cite{rivera-thorsenSunburstArcDirect2017}. 
\cite{sharonCosmicTelescopeThat2022} use a strong lensing model together with Chandra X-ray data to measure the projected cluster mass within 175 kpc $h^{-1}$ to be $2.05^{+0.01}_{-0.01}\times10^{14}\ M_\odot\ h^{-1}$. 
The mass models obtained consist of a total of 202 potentials, eight of which are large scale and optimized outside of the scaling relations.


\section{Methodology} \label{sec:methodology}

\subsection{Parametric lensing mass models} \label{subsec:parametric_models}

Mass maps derived from strong lensing measurements are typically constructed using parametric models, in which the mass distribution is described by a finite number of mass ``clumps'': the superposition of a few that represent the smooth large-scale cluster components, and multiple, smaller galaxy-scale components that are overlaid on top, around which GGSL events are detected. 
Each mass clump is described by a finite number of parameters contingent upon the specific choice of mass profile adopted. 
In addition, the mass distribution is generally assumed to be self-similar, with both large and small scale clumps being modeled using the same density profile.
Moreover, while current lensing measurements constrain the 2D projected mass enclosed within an aperture for chosen profiles, they are agnostic to the detailed spatial distribution of mass within that aperture.
On small scales, cluster mass modeling only provides integral constraints, namely, the total mass enclosed within an aperture.
Lensing-derived masses are therefore reliant on the modeling of the lens, and the choice of density profiles adopted can lead to differences in the estimated sub-halo masses \citep{meneghettiFrontierFieldsLens2017, minorRobustMassEstimator2017}. 

The most commonly used mass profile is the dual-Pseudo-Isothermal-Elliptical profile \citep[dPIE; or Pseudo-Isothermal Elliptical Mass Distribution, PIEMD; e.g., ][]{limousinConstrainingMassDistribution2005, kneibClusterLenses2011}.
This is the profile used by \texttt{LENSTOOL} in the publicly available mass reconstructions for the five clusters we consider in this study.
The dPIE projected mass distribution is given by \citep[e.g.,][]{eliasdottirWhereMatterMerging2007} \begin{equation}
    \Sigma(R) = \frac{3\pi\sigma_0^2}{2G}\frac{r_\mathrm{cut} + r_0}{r_\mathrm{cut}} \left( \sqrt{R^2 + r_0^2} - \sqrt{R^2 + r_\mathrm{cut}^2} + r_\mathrm{cut} - r_0 \right),
\end{equation} where $R$ is the 2D projected distance from the sub-halo center.
This profile is constrained by three parameters: the velocity dispersion ($\sigma_0$), core radius ($r_0$), and cut radius ($r_\mathrm{cut}$). 
The dPIE 3D profile behaves like an isothermal sphere ($R^{-2}$) between $r_0$ and $r_\mathrm{cut}$, but steepens to a $R^{-4}$ profile outside of the cut radius. 
It is a powerful lensing model due to its finite mass, its ability to reproduce a wide range of component sizes, and its analytically derivable lensing potential and derivatives. 
Furthermore, we can transform our coordinates to allow for non-spherical distributions via \begin{equation}\label{eq:ellip}
    R^2 = \left( \frac{x^2}{(1+\epsilon)^2} +  \frac{y^2}{(1-\epsilon)^2} \right).
\end{equation} 
The eccentricity $\epsilon$ and position angle of the ellipse represent two additional degrees of freedom. 
After accounting for sky position and redshift, we find that each model component is defined by eight parameters ($\sigma_0$, $r_0$, $r_\mathrm{cut}$, $\epsilon$, the position angle of the ellipse $\theta_\epsilon$, sky position $(x,y)$, and redshift $z$). 
In order to reduce the number of free parameters, $\sigma_0$ and $r_\mathrm{cut}$ for the small scale components are pegged to the luminosity of their galaxies using the scaling relations described in the M20sup.

The Navarro-Frenk-White \citep[NFW;][]{navarroUniversalDensityProfile1997} profile represents another parametric model that can be used to construct mass maps from lensing measurements.
This is the profile to which we rearrange the dPIE mass profiles obtained for the five clusters, in order to explore the effect on GGSL computations.
The NFW density profile has been shown to provide a good approximation for a universal mass profile for DM halos over a wide range of halo masses, from cluster scales to dwarf galaxy scales, and simulations tend to result in the production of halos and sub-halos that conform to the NFW profile. 
While there is debate regarding whether NFW-like profiles offer the best description for tidally stripped cluster sub-halos \citep[e.g.,][]{greenTidalEvolutionDark2019, greenTidalEvolutionDark2021}, since we are not following the dynamical evolution of sub-halos here, but rather studying snapshot lensing data for derived constraints, we adopt the truncated NFW profile, which we proceed to describe.

The NFW pofile is especially elegant in that it is completely specified by two parameters. These may be chosen to be $c$, the concentration parameter, and $r_{200}$, the radius inside which the density of halo is $200\rho_\mathrm{c}$, where $\rho_\mathrm{c}(z) = 3H^2(z)/8\pi G$ is the critical density for a flat universe.
The 3D density profile is then given by \citep[e.g., Eq. 1 in ][]{navarroUniversalDensityProfile1997}:
\begin{equation}
    \rho_\mathrm{NFW}\left(\frac{r}{r_\mathrm{s}}\right) = \frac{\delta_\mathrm{c}\rho_\mathrm{c}}{\frac{r}{r_\mathrm{s}}(1+\frac{r}{r_\mathrm{s}})^2}
\end{equation}
where $r$ is the 3D distance from the halo (or sub-halo) center, $r_\mathrm{s} = r_{200}/c$ is the so-called scale radius, and $\delta_\mathrm{c}$ is the (dimensionless) characteristic overdensity, given by 
\begin{equation} \label{eq:delta_c}
    \delta_\mathrm{c} = \frac{200}{3}\frac{c^3}{\ln(1+c) - c/(1+c)}.
\end{equation}
The characteristic feature of the NFW profile is that the power law index steepens from $-1$ to $-3$ near $r_\mathrm{s}$.
The overall shape, and crucially the concentration parameter $c$, offer reasonable fits to simulated data, as well as to observed profiles, including stacked lensing analyses.
In practice, the NFW is widely used for its simplicity and relative accuracy.

One drawback of the NFW profile is that the integrated halo mass is infinite, which is unphysical.
This is of special concern for cluster sub-halos, which are tidally stripped, and are therefore expected to have truncated mass profiles. 
\cite{baltzAnalyticModelsPlausible2009} put forth a truncated NFW (tNFW) model, which introduces a third parameter, $r_\mathrm{t}$, that corresponds to the tidal radius for tidally truncated halos.
The 3D density profile is given by:
\begin{equation}
    \rho_\mathrm{tNFW}\left(\frac{r}{r_\mathrm{s}}\right) = \rho_\mathrm{NFW}\left(\frac{r}{r_\mathrm{s}}\right) \times \begin{cases}
        \frac{\tau^2}{\tau^2 + \left(\frac{r}{r_\mathrm{s}}\right)^2} & \tau \geq 1 \\
        \frac{\tau^4}{\left(\tau^2 + \left(\frac{r}{r_\mathrm{s}}\right)^2\right)^2} & \tau < 1,
    \end{cases}
\end{equation}
where $\tau = r_\mathrm{t} / r_\mathrm{s}$.
The total integrated mass of a tNFW halo is finite.
For $r_\mathrm{t} \geq r_\mathrm{s}$, the 3D density profile drops off from $r^{-3}$ to $r^{-5}$ at $r_\mathrm{t}$, while for $r_\mathrm{t} < r_\mathrm{s}$ the profile goes to $r^{-5}$ at $r_\mathrm{t}$ and then further drops to $r^{-7}$ at $r_\mathrm{s}$.
As would be required, the tNFW profile reduces to an NFW for $\tau\to\infty$.

As described in Sec.~\ref{sec:ggsl_comparison_results}, we re-arrange the well-constrained mass enclosed within apertures that correspond to the truncation radius of the dPIE models provided by the best-fit lens model from \texttt{LENSTOOL}, and refit the enclosed mass derived from the best-fit mass models with best-fit tNFW profiles.

\subsection{Definition of the GGSL probability} \label{subsec:ggsl_def}

We now define the GGSL probability and other relevant quantities used in lensing computations.
Further details for all computations can be found in e.g., \cite{kneibClusterLenses2011} and \cite{meneghettiIntroductionGravitationalLensing2021}.
The two planes of interest for any lensing configuration are the \textit{source plane}, where the lensed background objects lie, and the \textit{lens plane}, where the foreground mass and hence the lensing potential resides and in which the lensed images appear.
We adopt the thin lens approximation, in which the source and lens planes are approximated as 2D planes, as it is only projected quantities that are relevant for calculating the lensing signal.
This approximation is valid since the physical extent of the lenses of interest (i.e., galaxies and clusters) are significantly smaller than the distances from the observer to the lens and source.
Structures in the source plane are mapped onto the lens plane using the deflection induced by the projected surface mass density $\Sigma(\mathbf{\theta})$, where $\mathbf{\theta}$ is the 2D vector position on the lens plane. 

The deflection of light and the distortions it produces can be thought of as effecting a mathematical transformation that occurs between the source plane and lens plane and is characterized by two quantities: the \textit{convergence} ($\kappa$) that quantifies the isotropic (radial) focusing of light and affects the apparent size and brightness of a galaxy, but not its shape; and the \textit{shear} ($\gamma$), a tensor that quantifies the stretching or compression of light rays tangential to the lensing object's center, causing elliptical distortions to the background galaxy image.
A positive (negative) convergence magnifies (reduces) the apparent size and brightness of the galaxy; the orientation of the shear indicates the axis along which the galaxy image is stretched.
$\kappa(\mathbf{\theta})$ is calculated from $\Sigma(\mathbf{\theta})$ and the components of $\gamma(\mathbf{\theta})$ are calculated from $\kappa(\mathbf{\theta})$ and partial derivatives of the \textit{deflection angle} $\alpha(\mathbf{\theta})$, which is the angle by which light rays deviate from a straight line.
From $\kappa(\mathbf{\theta})$ and $\gamma(\mathbf{\theta})$, we can define where in the lens plane $\Sigma(\mathbf{\theta})$ exceeds a critical value ($\Sigma_\mathrm{crit}$), thus generating \textit{critical lines} at the contours defined by \citep[see e.g., Eq.~16 of][]{kneibClusterLenses2011}\begin{equation*}
    \kappa(\mathbf{\theta}) \pm \gamma(\mathbf{\theta}) = 1,
\end{equation*}
where $\gamma(\mathbf{\theta})$ is the norm of the shear.
The $+$ corresponds to \textit{tangential} critical lines, where background sources become infinitely stretched (resulting in arcs and Einstein rings) and the $-$ to \textit{radial} critical lines, where background sources become infinitely magnified. 
Since observations of strong lensing by the galaxies seen in our cluster samples are expected to have negligible radial critical lines (see M20sup), we focus only on the tangential critical lines\begin{equation} \label{eq:critical_lines}
    \mathbf{\theta_t}: \kappa(\mathbf{\theta}) + \gamma(\mathbf{\theta}) = 1.
\end{equation}
Ultimately, any particular $\mathbf{\theta_t}$ curve produced by the small scale clumps is due to the entirety of $\Sigma(\mathbf{\theta})$ in the lens plane including the boosting provided by the underlying smooth gravitational potential of the larger scale clumps. 
Critical lines can be assigned a size according their enclosed area $A_\mathrm{enc}$, which can serve to roughly associate the critical line to the mass of the ``subclump'' that gave rise to it. 
From $A_\mathrm{enc}$ we can further define the \textit{effective Einstein radius} of the critical line \begin{equation}
    \theta_\mathrm{E} = \sqrt{\frac{A_\mathrm{enc}}{\pi}}.
\end{equation}

We wish to distinguish between the critical lines that are produced by the larger cluster-scale mass distribution, and those that are produced by the smaller scale individual cluster sub-halos (and thus are attributable to GGSL).
To do so, we adopt the convention of \cite{meneghettiProbabilityGalaxygalaxyStrong2022} wherein critical lines with $\theta_\mathrm{E}>5''$ are considered \textit{primary}, those with $0.5''<\theta_\mathrm{E}<3''$ are considered \textit{secondary}, and all others are excluded for the purpose of this small-scale analysis.
$\theta_\mathrm{E}=0.5''$ corresponds to a physical radius of $\sim2.3$ kpc and a mass of $\sim10^{10}\ M_\odot$, assuming a point mass lens at $z=0.3$.
It is the secondary critical lines produced by the smaller scale clumps---cluster sub-halos---that we define as giving rise to GGSL events.
Our criterion for secondary critical lines is motivated by the ability to effectively compare observational results with simulations: anything less than $0.5''$ cannot be resolved by the numerical simulations considered for comparison in this work.
Moreover, this is $\sim5$ times larger than the resolution of HST images.
The $3''<\theta_\mathrm{E}<5''$ regime, which implicates group-scale masses, are ones that we exclude by definition of the secondary critical lines given our intent to focus on properties of individual cluster galaxy sub-halos.
Simulations, meanwhile, tend to have significant lensing events due to critical lines of this size indicating the presence of in-falling groups in these massive assembling clusters \citep[][]{meneghettiProbabilityGalaxygalaxyStrong2022, ragagninGalaxiesCentralRegions2022, meneghettiPersistentExcessGalaxygalaxy2023}.

Any critical line $\mathbf{\theta_t}$ can be mapped to a corresponding curve in the lens plane, called a \textit{caustic line} \citep[e.g., Eq.~3.97 of][]{meneghettiIntroductionGravitationalLensing2021}: \begin{equation}
    \mathbf{\beta_t} = \mathbf{\theta_t} - \alpha(\mathbf{\theta_t}).
\end{equation}
The total area enclosed by the secondary caustics $\mathbf{\beta_t}$ in the source plane is defined to be the \textit{GGSL cross-section}, $\sigma_\mathrm{GGSL}$ (see the right-hand panel of Fig.~\ref{fig:mass_map_caustics}).

Placing the background sources and hence the source plane at different redshifts $z$ results in different values of $\alpha(\mathbf{\theta})$ and therefore different curves $\mathbf{\theta_t}$ and different caustics $\mathbf{\beta_t}$, so in general, $\sigma_\mathrm{GGSL} = \sigma_\mathrm{GGSL}(z)$. 
From the GGSL cross-section, we can further define the \textit{GGSL probability}, $P_\mathrm{GGSL}$, as the ratio of the area enclosed by the secondary caustics to the area in the source plane that maps to the entire FoV in the lens plane, $A_s$:\begin{equation} \label{eq:p_ggsl}
    P_\mathrm{GGSL}(z) = \frac{\sigma_\mathrm{GGSL}(z)}{A_s(z)}.
\end{equation}
This is the quantity that was shown by M20 to be discrepant between cluster lensing data and \lcdm\ simulations.

\subsection{Computational procedure} \label{subsec:procedure}
\begin{figure*}
    \centering
    \includegraphics[width=\textwidth]{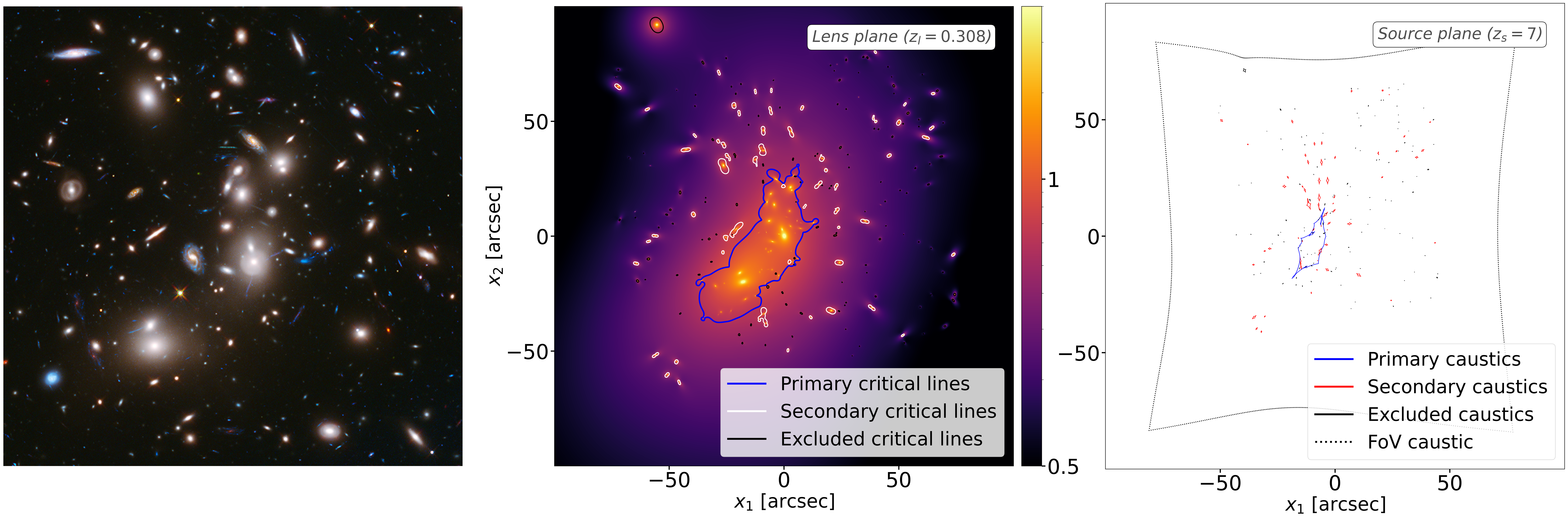}
    \caption{Our GGSL probability calculation pipeline illustrated with the HFF cluster Abell 2744. \textbf{Left:} The HFF image of Abell 2744; \textbf{Center:} The projected 2D mass density reconstruction from parametric dPIE lensing mass model. The colorbar indicates the scale of $\kappa$, the convergence, represented by the heatmap. Contours show the critical lines calculated with the \texttt{lenstronomy} software package for a source plane at $z=7$. The central blue critical line is the only primary critical line in this case and the secondary critical lines are shown in white. In black we show the critical lines that are excluded as described in the text. \textbf{Right:} The corresponding primary, secondary, and excluded caustics in the source plane. The FoV caustic is the caustic associated with the FoV of the image plane. The area enclosed by the secondary caustics defines the GGSL cross-section, and the ratio of GGSL probability to the total area enclosed by the FoV caustic defines the GGSL probability. Image credit: Abell 2744 at \url{https://hubblesite.org/}.
 \label{fig:mass_map_caustics}}
\end{figure*}
As stated above, magnification maps obtained from observations of clusters assume mass-model priors, and in this paper we explore how altering the DM distribution model within the sub-halos impacts the GGSL probability.
We compute $P_\mathrm{GGSL}(z)$ for each of the five clusters in our sample for both the dPIE and tNFW models. 
For both cases, we do so using a method independent of the one used in M20 by utilizing the \texttt{lenstronomy}, a package that implements lensing calculations in Python \citep{birrerLenstronomyMultipurposeGravitational2018, birrerLenstronomyIIGravitational2021}.\footnote{\url{https://github.com/lenstronomy}} 
The steps of this procedure to compute GGSL from the \texttt{lenstronomy} package are outlined below:
\begin{itemize}
    \item For each cluster, we take a list of potentials from the best-fit lensing model (either the \texttt{LENSTOOL} dPIE or the re-fitted tNFW models) that includes the properties of each potential, its position in the lens plane, and other relevant parameters for the dPIE or tNFW.
    \item Consistent with the FoV used in M20, we impose a $200''\times 200''$ FoV for all clusters to normalize $P_\mathrm{GGSL}$, which cuts some potentials out of the computation.  This ensures that the normalization of $P_\mathrm{GGSL}$ is the same in our study as in M20.  See Eq.~\ref{eq:p_ggsl}.
    \item We then pass the list of potentials to \texttt{lenstronomy}, along with some source plane redshift $z_i$, and we use its functionality to construct maps of $\Sigma(\mathbf{\theta})$ and $\alpha(\mathbf{\theta})$, as well as the relevant derivatives for calculating $\kappa(\mathbf{\theta})$ and $\gamma(\mathbf{\theta})$.
    \item Using those results, we use Eq.~\ref{eq:critical_lines} to return lists of image pixels that make up the critical lines.
    \item We calculate the equivalent Einstein radius for each critical line, and label each critical line as primary or secondary as shown in 
    Fig.~\ref{fig:mass_map_caustics}, central panel.
    \item We use \texttt{lenstronomy} to map the secondary critical lines to their corresponding caustics in the source plane (Fig.~\ref{fig:mass_map_caustics}, right panel).
    \item The area enclosed by the secondary caustics is calculated and summed to compute $P_\mathrm{GGSL}(z_i)$.
    \item We iterate this procedure over different values of $z_i$ to calculate $P_\mathrm{GGSL}(z)$.
\end{itemize}


\section{GGSL results with dPIE and tNFW fits} \label{sec:ggsl_comparison_results}

\begin{figure*}
    \gridline{
        \fig{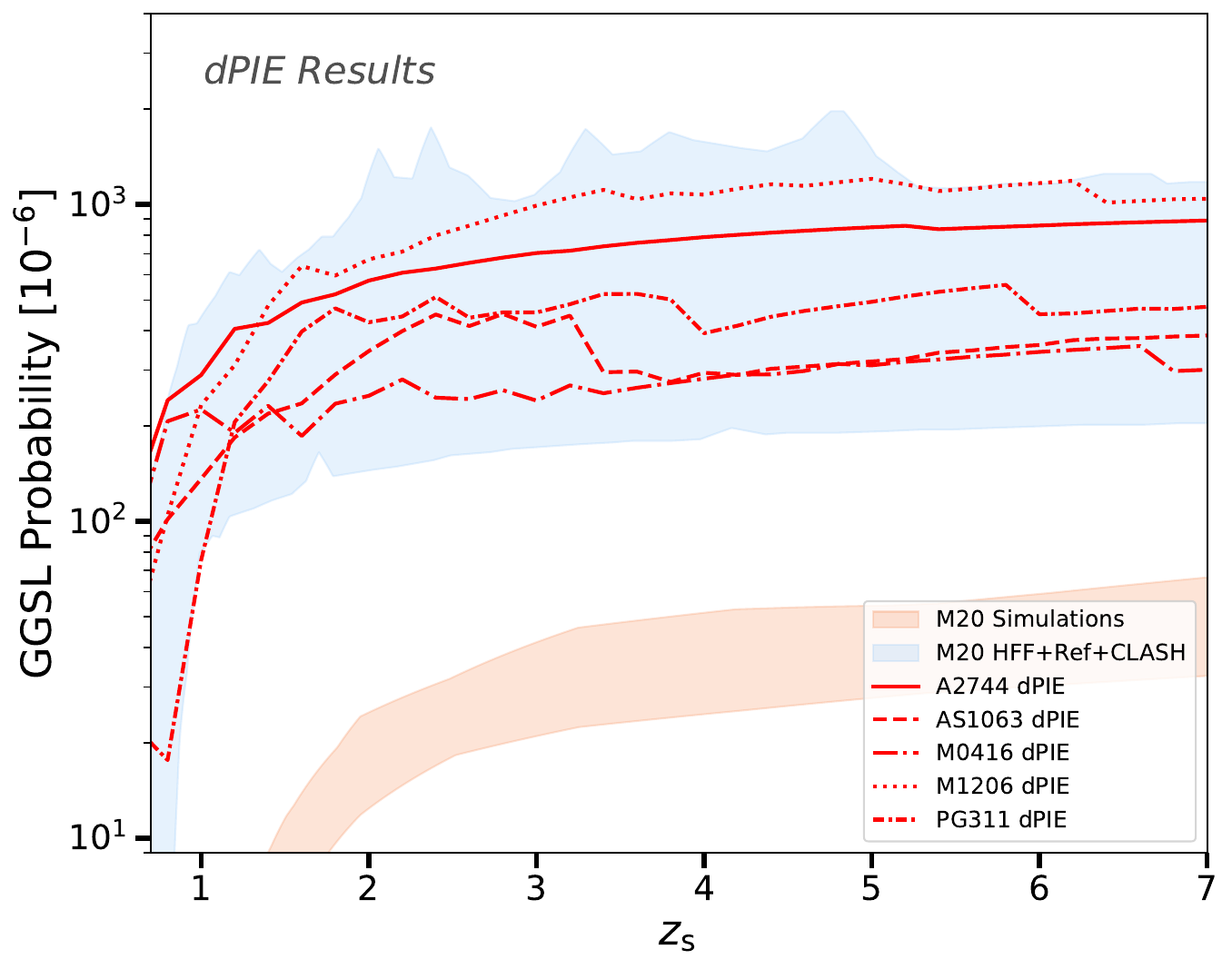}{0.49\textwidth}{(a)}
        \fig{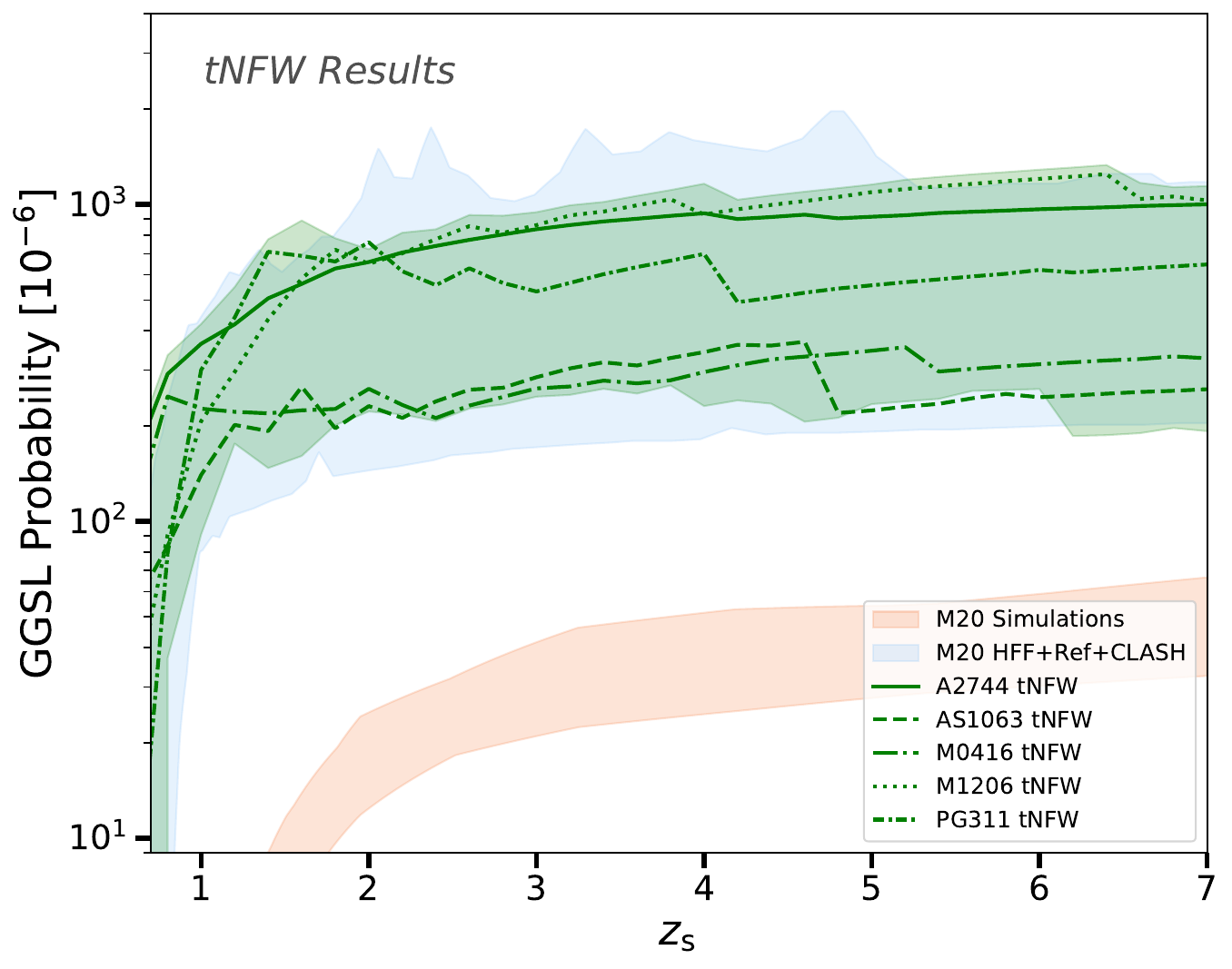}{0.49\textwidth}{(b)}
    }
    \caption{$P_\mathrm{GGSL}$ as a function of source redshift for the five clusters, shown for both the dPIE and tNFW mass models, compared with the results of M20. The ``M20 HFF+Ref+CLASH'' band is a composite of the three samples studied in M20, combined for simplicity, and is included to guide the eye. The orange band represents the simulation sample of M20. (a) dPIE results for the five clusters in our study recovers the result of M20. (b) Redistributing the mass within $r_\mathrm{cut}$ as a tNFW profile does not significantly change the GGSL probability. Confidence intervals shown are based on 95\% confidence in the tNFW concentration parameter. \label{fig:side_by_side_results}}
\end{figure*}

Following the procedure outlined in Sec.~\ref{subsec:procedure}, we use the publicly available dPIE fits from \texttt{LENSTOOL} to independently calculate $P_\mathrm{GGSL}(z)$ for each cluster using the \texttt{lenstronomy} software package.
While \texttt{LENSTOOL} and \texttt{lenstronomy} perform similar computations,\footnote{There are slight differences in, e.g., how ellipticity is calculated.  \texttt{lestronomy} performs the coordinate transformation described in Eq.~\ref{eq:ellip} on the lensing potential field, while \texttt{LENSTOOL} performs the transformation directly on the mass density field.  This can result in slight discrepancies for small ellipticities, such as those considered in this study.} \texttt{lenstronomy} is a more flexible and user-friendly package, since it is written for Python, and it also offers many more parametric models for the lensing potentials (for example, tNFW is not offered by \texttt{LENSTOOL}).
In M20, GGSL was computed using the lensing mass model derived from \texttt{LENSTOOL} in which the sub-halos are modeled with a dPIE.
Here, we first re-calculate GGSL from the lensing mass models with dPIE sub-halos using \texttt{lenstronomy} and find that our results are in agreement with those reported in M20 for a sub-sample of their clusters that now includes an additional massive cluster lens that was found via a completely different selection technique, its SZ emission.

Our results are shown in Fig.~\ref{fig:side_by_side_results}.
In both panels, the blue band labeled "M20 HFF+Ref+CLASH" is a composite of the three observed samples studied in M20, and the orange band is their simulation set (both are adapted from Fig.~3 of M20).
We briefly describe each of those data sets here; more details can be found in M20sup.
The eleven HST clusters used by M20 for their observational data set consist of four HFF clusters, four clusters from the CLASH ``gold'' sample, and three additional clusters with extremely well-constrained mass models (the ``reference sample'').
The GGSL probability was measured as a function of redshift for each of the clusters.
M20 also performed the same analysis for 25 simulated clusters, matched in mass, morphology, redshift, and mass concentration to the observed samples.
Cluster halos from a parent DM-only simulation with $1024^3$ particles were identified in a periodic low resolution simulation box with co-moving size of 1 $h^{-1}$ Gpc using a standard friends-of-friends algorithm.
The simulation is a flat \lcdm\ universe with cosmological parameters $h=0.72$, $\sigma_8=0.8$, $\Omega_{m,0}=0.24$, $\Omega_{b,0}=0.04$, and an adopted primordial power spectrum of $P(k)\propto k^{0.96}$.
The clusters were then re-simulated with baryonic physics using the TreePM-SPH code GADGET-3 several times, to include different sub-grid physics models.
These incorporated different prescriptions for gas cooling, star formation, and energy feedback from supernovae and active galactic nuclei (AGN; some variants even turned AGN feedback off entirely).
The mass of DM particles in these simulations was $8.47\times10^8\ h^{-1}\ M_\odot$ and that of gas particles was $1.53\times10^8\ h^{-1}\ M_\odot$.
To identify potential inaccuracies due to limited resolution, a sub-sample of clusters were re-simulated with a factor of ten increased resolution with no change to the results.
M20 also verified their results by producing mock realizations of the clusters using the semi-analytic code MOKA \citep{giocoliMokaNewTool2012}.
To compute the GGSL probability in the simulated clusters, they smoothed the particles over the cluster volume and employed ray-tracing to compute deflection angle maps from their projected mass distributions.

As shown in Fig.~\ref{fig:side_by_side_results}(a), we have recovered the results of M20 for the dPIE case, and our independent computation of the GGSL probability is also discrepant with their \lcdm\ simulation results.
We then investigate the effect of redistributing the mass enclosed within cluster sub-halos using tNFW profiles.
Recall that lensing mass models only offer an integral constraint on the inferred mass enclosed within a fixed aperture for sub-halos, and are agnostic to how the mass is distributed radially inside sub-halos.
The tNFW profile has three degrees of freedom, which can be represented by $r_{200}$, $c$, and $\tau$.
We require that $M_\mathrm{cut}$, the mass enclosed at $r_\mathrm{cut}$ of the dPIE profile, remains constant, which reduces the number of degrees of freedom to two.
Note that even if we allow all of $M_\mathrm{cut}$ to contract into a single point mass, corresponding to a dPIE inner slope of -2.99, the computed $P_\mathrm{GGSL}$ lies within the band derived for the observed lensing samples by M20.
We constrain $\tau$ and $r_\mathrm{s}$ using a least squares fit to the remaining dPIE profile between $r_0$ and $r_\mathrm{cut}$, and then determine $c$ using those best-fit values and the fixed data point at $(r_\mathrm{cut}, M_\mathrm{cut})$.
We estimate 95\% confidence intervals with a Monte-Carlo method, using 500 iterations for each sub-halo with random selections of data points for the least squares fit. 
See Appendix~\ref{sec:app_fitting} for a detailed description of the fitting procedure and error estimation.
We retain the largest-scale potentials as dPIE profiles, since we are interested in how these changes specifically affect GGSL, while the background cluster density remains similar.
We passed these new tNFW models to \texttt{lenstronomy} to calculate $P_\mathrm{GGSL}(z)$, and the results are shown in Fig.~\ref{fig:side_by_side_results}(b).
Note that there is some cluster to cluster variation in the computed GGSL, with the more massive cluster lenses having a higher GGSL probability.

We have thus demonstrated that changing the mass model for the sub-halos does not resolve the reported GGSL discrepancy with \lcdm\ simulations.


\section{Effect of inclusion of baryonic contributions to halo substructure on GGSL} \label{sec:baryons}

\begin{figure}
    \centering
    \includegraphics[width=\textwidth]{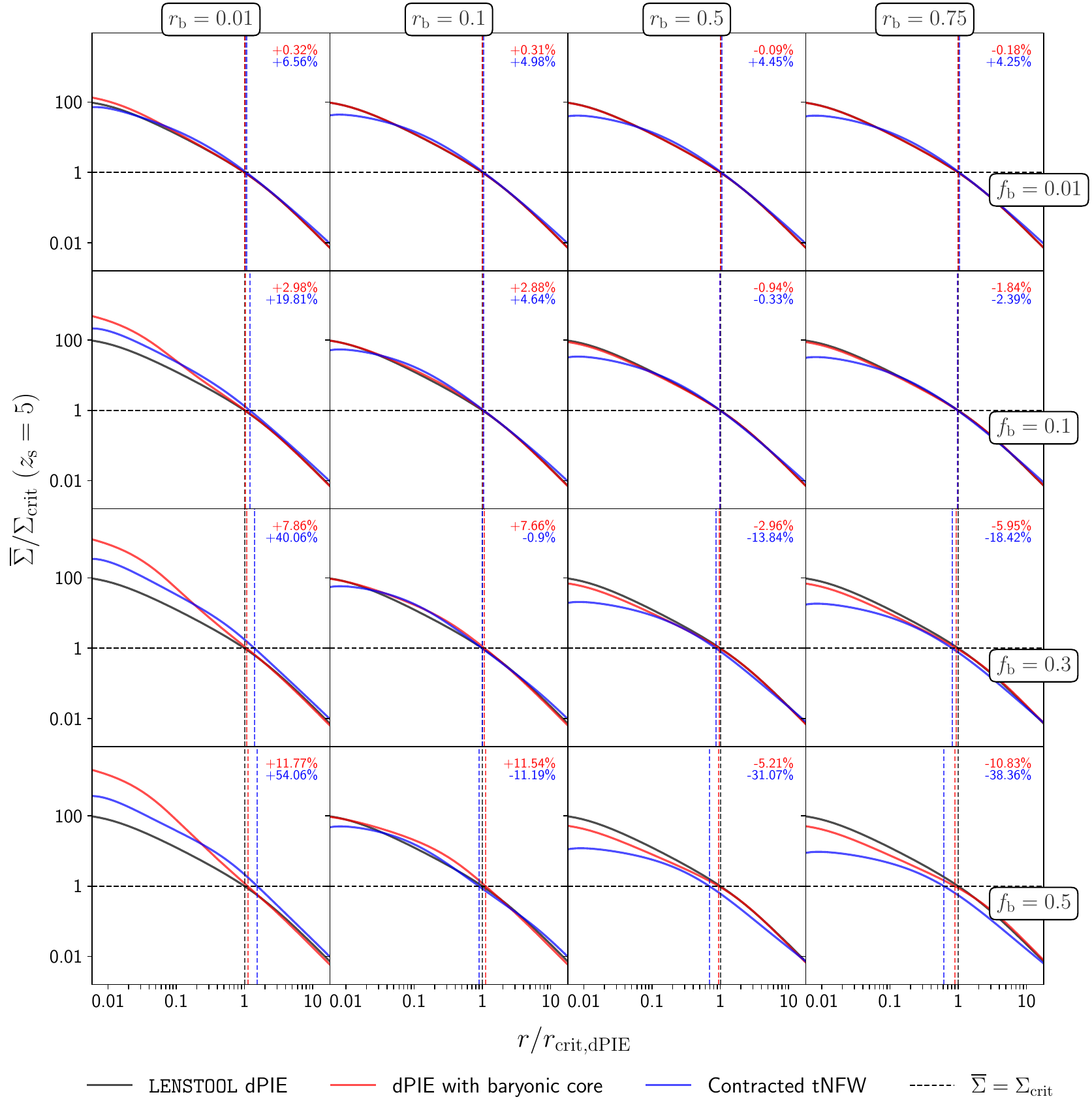}
    \caption{Each plot shows the Einstein radii for three cases of inner mass distribution for a representative sub-halo from the HFF cluster M1206. The Einstein radius coincides with the critical line of an isolated sub-halo, and is correlated with GGSL probability. The numbers in the top-right of each panel are the percent change of $r_\mathrm{E}$ relative to that of dPIE profile. 
    \textbf{Black:} the \texttt{LENSTOOL} dPIE profile, to which all other profiles are normalized; 
    \textcolor{red}{\textbf{Red:}} the dPIE profile modified to include an exponential baryonic component such that $f_\mathrm{b}$ of the mass within $r_\mathrm{cut}$ is in the baryonic core, and with scale length $r_\mathrm{d} = r_\mathrm{b}\times r_\mathrm{cut}$;
    \textcolor{blue}{\textbf{Blue:}} our best-fit tNFW profile after an adiabatic contraction such that the final baryonic density profile is the exponential core described above. We look at four different values of baryon mass fraction ($f_\mathrm{b}$) and baryon scale radius fraction relative to $r_\mathrm{cut}$ ($r_\mathrm{b}$). Even in the most extreme cases, we do not see significant enough changes to $r_\mathrm{crit}$ to account for the GGSL disparity reported in M20.} 
    \label{fig:r_crit_grid}
\end{figure}

As shown above, redistributing the matter within each sub-halo into an tNFW profile accounts for the total mass.
However, virtually all halos that contribute to GGSL host luminous, baryonic galaxies.
Baryons affect the concentration of their dark matter hosts in a number of ways: they can steepen the inner slope due to adiabatic contraction \citep{gnedinResponseDarkMatter2004}, but also soften the slope due to dynamical friction \citep{pettsSemianalyticDynamicalFriction2015, banikDynamicalFrictionBuoyancy2022}, supernova feedback \citep{pontzenHowSupernovaFeedback2012}, and 3-body interactions.
We focus our attention here on adiabatic contraction, since we have already probed the cored dPIE profiles, and the other effects will make the tNFW develop a core.
Exploring the effects of adiabatic contraction allows us to explore a greater range of even more cuspy profiles beyond the tNFW that are still consistent with \lcdm.

We focus on a single representative sub-halo from M1206 (\verb|gal_id 4229|) and modify its profile in two ways.
We chose to analyze a sub-halo from M1206 because that cluster has the largest measured $P_\mathrm{GGSL}$ from our sample, as seen in Fig.~\ref{fig:side_by_side_results}.
\verb|gal_id 4229| was optimized by \texttt{LENSTOOL} using the scaling relations for M1206, and does not have any particular distinguishing feature that would make it an outlier as far as GGSL, such as ellipticity or a peripheral position within the cluster.
First, we explore the effect of adding an exponential core, meant to represent the baryons, to the sub-halo's \texttt{LENSTOOL} dPIE profile.
The dPIE profile is then renormalized in order to conserve $M_\mathrm{cut}$, which is the observational constraint obtained from the best fit lensing mass models. 
Second, we apply an adiabatic contraction to our tNFW fits, with the same exponential 3D density profile as the final baryonic profile representing the hosted galaxy.
We adopt the the adiabatic contraction model of \cite{gnedinResponseDarkMatter2004}, implemented in the publicly available adiabatic contraction code \texttt{Contra}.
Note that we conserve mass within $r_\mathrm{cut}$ provided by lensing mass models to within $\sim10\%$.

For a baryonic component of total mass $M_0$, we use an exponential 3D density profile given by:
\begin{align}
    \rho(r) &= \frac{M_0}{4\pi r_d^2 r}e^{-r/r_\mathrm{d}}
\end{align} which yields the enclosed mass profile \begin{align}
    M_\mathrm{3D}(r) &= M_0\left[ 1 - \left(1 + \frac{r}{r_\mathrm{d}}\right)e^{-r/r_\mathrm{d}} \right].
\end{align}  Using an Abel transform, we can derive the projected 2D profiles \begin{align}
    \Sigma(R) &= \frac{M_0}{2\pi r_\mathrm{d}^2}K_0\left(\frac{R}{r_\mathrm{d}}\right); & M_\mathrm{2D}(R) &= M_0\left[ 1 - \frac{R}{r_\mathrm{d}}K_1\left(\frac{R}{r_\mathrm{d}}\right) \right],
\end{align} where $K_n$ is the $n^\mathrm{th}$ order modified Bessel function of the second kind.
We set $M_0$ by satisfying $M_\mathrm{2D}(r_\mathrm{cut}) = f_\mathrm{b}M_\mathrm{cut}$, where $M_\mathrm{cut}$ is the mass enclosed within $r_\mathrm{cut}$ in the dPIE model of the sub-halo and $f_\mathrm{b}$ represents the baryon mass fraction.
The baryonic scale radius $r_\mathrm{d}$ is given by $r_\mathrm{d} = r_\mathrm{b}r_\mathrm{cut}$.

In order to quantify the effect of these changes on the GGSL probability, we calculate the Einstein radius $r_\mathrm{crit}$, which defines the critical line for an isolated halo, for each profile.
Since the GGSL cross section is defined as the area enclosed by secondary caustic in the source plane, which are mappings of secondary critical lines, $r_\mathrm{crit}$ is a proxy for GGSL probability. 
See Appendix~\ref{sec:app_einstein_radius} for a demonstration of this relationship. We can calculate $r_\mathrm{crit}$ for each profile using the relation $\overline{\Sigma}(r_\mathrm{crit}) = \Sigma_\mathrm{crit}$ \citep[e.g., Eq. 5.34 in][]{meneghettiIntroductionGravitationalLensing2021}, where $\overline{\Sigma}$ is the average surface density and $\Sigma_\mathrm{crit}$ is the critical lensing surface density, which is a function of the lens plane redshift and source plane redshift alone.
We perform this procedure for a range of baryon mass fractions ($f_\mathrm{b}$) and baryonic disk scale radius fraction ($r_\mathrm{b}$).
The stellar mass-halo relation for local low mass galaxies (up to $M_*=10^9$) derived by \cite{zaritskyPhotometricMassEstimation2023}, and that of \cite{girelliStellartohaloMassRelation2020} for higher mass halos at $0\leq z\leq 4$, both predict $5\times10^{-4}\lesssim f_\mathrm{b} \lesssim 10^{-1}$ for the range of sub-halo masses in M1206. We probe even higher values in our analysis to also explore the potential effects of the even more extreme cluster environment on lensing properties.

The results of our analysis are shown in Fig.~\ref{fig:r_crit_grid}.
In most cases, the change to $r_\mathrm{crit}$ is imperceptible, while in the most extreme cases, the percent change to $r_\mathrm{E}$ reaches $\sim54\%$ on the positive end and $\sim-38\%$ on the negative end.
For a sub-halo with a $r_\mathrm{E}\sim0.9''$ at $z_\mathrm{s}=5.0$, these most extreme values correspond to changes in $P_\mathrm{GGSL}$ by factors of approximately 2.5 and 0.35, respectively.
This would still leave three of the five curves in Fig.~\ref{fig:side_by_side_results}(a) within the M20 band (at least for the negative changes---extreme positive changes would further exacerbate the tension), while M0416 and AS1063 would still be more than two times off from the $+95\%$ confidence interval of the ``M20 simulations'' band.


\section{Conclusions} \label{sec:conc}

The GGSL discrepancy in cluster lenses motivates the study of the concentration-mass (c-M) relation for cluster sub-halos, which might provide new insights into their inner density profiles.
\cite{navarroUniversalDensityProfile1997} note that since less massive systems collapse at higher redshifts, they are expected to have higher concentration parameters.
This relation for DM halos has been modeled and studied extensively, primarily through N-body simulations \citep[e.g.,][]{bullockProfilesDarkHaloes2001, ekePowerSpectrumDependence2001, wechslerConcentrationsDarkHalos2002, zhaoGrowthStructureDark2003, maccioConcentrationSpinShape2008, zhaoAccurateUniversalModels2009, diemerUniversalModelHalo2015, ludlowMassConcentrationRedshift2016, gilmanWarmDarkMatter2020}.
Observationally, it has been demonstrated that strong lensing measurements are a powerful probe of the c-M relation \citep[e.g.,][]{vegettiDensityProfileDark2014, vegettiInferenceColdDark2014, minorUnexpectedHighConcentration2021, gilmanConstraintsMassConcentration2020}.
However, a systematic study of the observed c-M relation among cluster sub-halos remains to be done.
As mentioned above, \cite{minorUnexpectedHighConcentration2021} calculated a higher-than-expected concentration for the halo of the field galaxy SDSSJ0946+1006, and the higher efficiency of GGSL in observed clusters suggests that perhaps cluster sub-halos are also over-concentrated.

In this work, we address the GGSL tension as reported in M20, in which observed cluster lenses exhibit more efficient GGSL than their simulated counterparts.
We do so within the \lcdm\ cosmological paradigm, and not through exploration of alternative DM models, by leveraging the fact that lensing measurements only make integral mass constraints on the projected mass maps of cluster lenses.
We are thus afforded the freedom to redistribute the mass of cluster sub-halos within the observed $r_\mathrm{cut}$, within which the total enclosed mass is constrained.
Four of the clusters we study are part of the M20 study, while PG311 was part of later studies \citep{meneghettiProbabilityGalaxygalaxyStrong2022, meneghettiPersistentExcessGalaxygalaxy2023}, and is included because it was selected using a different criterion.
We first recover the results of M20, which show that using the cored dPIE profile as the mass model, the $P_\mathrm{GGSL}$ for each cluster is an order of magnitude higher than that of \lcdm-simulated counterparts.
This serves both to demonstrate the robustness of the M20 results, and to validate our $P_\mathrm{GGSL}$ computation methods, which are distinct from those of M20 in that they use the \texttt{lenstool} Python package.
We then refit the sub-halos of each cluster to the ``cuspy'' tNFW profile, while conserving mass within $r_\mathrm{cut}$, which yielded a comparable GGSL probability for the clusters.
Finally, we further explored the effects of modifying the DM sub-halo profiles with inclusion of a separate contribution from the baryonic components, accounting for the effects of adiabatic contraction.
In all cases, even with the most extreme assumed parameters for the baryonic disk component, we found that the lensing properties do not change significantly.
Thus, we demonstrate with detailed and careful analyses that the redistribution of the mass within cluster sub-halos, be it the overall density profile or the detailed distribution of baryons in the inner regions of the sub-halo, does not significantly change the GGSL probability that is obtained for lensing clusters.

While we have shown that a variety of profiles that are consistent with \lcdm\ yield similar $P_\mathrm{GGSL}$ results, GGSL computations do rely heavily on consistent and well-motivated definitions of primary and secondary critical lines, as discussed above (Sec.~\ref{sec:methodology}).
These are typically related to their effective Einstein radii, $\theta_\mathrm{E}$.
The results reported in \cite{robertsonGalaxyGalaxyStrong2021} are due to their inclusion of all critical lines with $0.5''\leq\theta_\mathrm{E}\leq5.0''$ as secondary critical lines, which is inconsistent with the smaller scale cut of $0.5''\leq\theta_\mathrm{E}\leq3.0''$ made in M20. \cite{meneghettiProbabilityGalaxygalaxyStrong2022} clearly demonstrate that the contribution from individual sub-halos to GGSL is dominated by those with $\theta_\mathrm{E}\leq3.0''$. In fact, Einstein radii of $3.0''<\theta_\mathrm{E}<5.0''$ correspond to the group scale sub-halos with masses $\sim 10^{13}\ M_\odot$.
From Fig.~1 in \cite{robertsonGalaxyGalaxyStrong2021}, it is clearly seen that it is the inclusion of the contribution of the critical line of the in-falling group in the lower right hand corner that boosts the GGSL signal.
Their inclusion boosts the computation of the GGSL presented by \cite{robertsonGalaxyGalaxyStrong2021} leading to their claim of having resolved the discrepancy.  
\cite{meneghettiProbabilityGalaxygalaxyStrong2022} exclude critical lines with $3''<\theta_\mathrm{E}<5''$, as we have done here.
As noted above, we reiterate that the increase in mass resolution alone does not enhance GGSL to account for the order of magnitude discrepancy reported in M20.

Ongoing surveys by the ESA Euclid mission \citep{sartorisNextGenerationCosmology2016} and the planned ones for the upcoming NASA Nancy Grace Roman Space Telescope mission \citep{spergelWFIRST2WhatEvery2013} are likely to provide high quality data for many more cluster lenses, dramatically increasing the sample size available for GGSL analyses \citep[e.g.,][]{mantzFutureLandscapeHighRedshift2019}.
This high resolution imaging coupled with spectroscopic follow-up will permit building accurate lensing models for these larger cluster lens samples as well.
Additionally, many computational groups are deeply invested in improving the implementation of baryonic physics and feedback processes in CDM simulations.
This will provide more sophisticated simulated clusters for comparison with observed cluster lenses.

Here we have examined a possible solution to the GGSL discrepancy within the \lcdm\ framework, and conclude that we are unable to resolve it by modifying the lens mass model through redistributing the matter within the sub-halo apertures, neither by invoking cusps nor the contribution of baryons in the inner regions of sub-halos. 
As noted in M20 in the first reporting of the GGSL discrepancy, two potential solutions within \lcdm\ could be held to account. 
One is that because we get integral constraints with the lensing models, there is freedom in how mass is distributed the within sub-halos. 
The second is the numerical resolution of simulations. 
In \cite{ragagninGalaxiesCentralRegions2022} and \cite{meneghettiProbabilityGalaxygalaxyStrong2022}, it was shown that numerical resolution alone cannot account for the observed tension.
Here we demonstrate that redistribution of mass via physical processes available within the context of the CDM paradigm also cannot account for the unexpectedly efficient observed GGSL.
Our conclusion suggests that exploration of alternative DM models and their impact on the lensing properties of cluster sub-halos is warranted.


\begin{acknowledgments}
We gratefully acknowledge: Frank Van den Bosch for his valuable feedback; Daniel A. Gilman for his consultation with regard to using the \texttt{lenstronomy} software package. P.N. acknowledges support from DOE grant \#DE-SC0017660.
This work utilizes gravitational lensing models produced by PIs Bradač, Natarajan \& Kneib (CATS), Merten \& Zitrin, Sharon, Williams, Keeton, Bernstein and Diego, and the GLAFIC group. This lens modeling was partially funded by the HST Frontier Fields program conducted by STScI. STScI is operated by the Association of Universities for Research in Astronomy, Inc. under NASA contract NAS 5-26555. The lens models were obtained from the Mikulski Archive for Space Telescopes (MAST).
\end{acknowledgments}

\software{
    Astropy \citep{astropycollaborationAstropyCommunityPython2013, astropycollaborationAstropyProjectBuilding2018, astropycollaborationAstropyProjectSustaining2022},
    Contra \citep{gnedinResponseDarkMatter2004},
    Lenstronomy
    \citep{birrerLenstronomyMultipurposeGravitational2018, birrerLenstronomyIIGravitational2021}
    }


\appendix

\section{tNFW fitting}
\label{sec:app_fitting}

\begin{figure}
    \centering
    \includegraphics[width=0.5\textwidth]{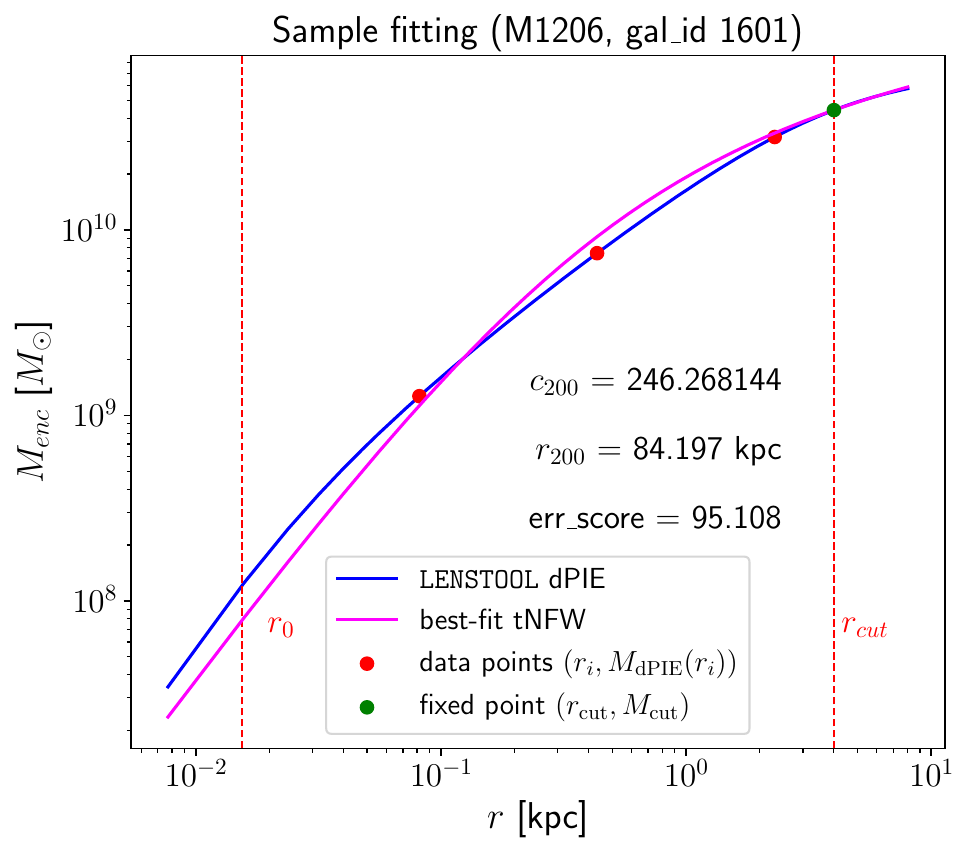}
    \caption{Example of one tNFW fitting of one PIEMD potential from the cluster M1206.  The green dot is the point $(r_\mathrm{cut}, M_\mathrm{cut})$, which is fixed.  The two remaining free parameters of are tuned to minimize the least squared error between the magenta curve and the red points.} 
    \label{fig:example_fit}
\end{figure}

We fit the publicly available \texttt{LENSTOOL} dPIE potentials, which are described by projected 2D mass profiles, to projected 2D tNFW mass profiles given by \cite{baltzAnalyticModelsPlausible2009}: \begin{equation} \label{eq:m_tNFW}
    M_\mathrm{2D}\left( \frac{R}{r_\mathrm{s}} ; c, \tau, r_\mathrm{s} \right) = 4\pi r_\mathrm{s}^3\rho_\mathrm{c} \times \delta_\mathrm{c}(c) \times \beta\left(\frac{R}{r_\mathrm{s}};\tau\right)
\end{equation} with \begin{align*}
    \beta\left(\frac{R}{r_\mathrm{s}};\tau\right) &= \begin{cases}
        \frac{\tau^2}{(\tau^2 + 1)^2} \left\{ \left[ \tau^2 + 1 + 2\left(\left( \frac{R}{r_\mathrm{s}} \right)^2 - 1\right) \right]  f\left(\frac{R}{r_\mathrm{s}}\right) \right. \\
        \left. \hspace{35pt} + \tau\pi + \left( \tau^2 - 1 \right)\ln\tau + \sqrt{\tau^2 + \left( \frac{R}{r_\mathrm{s}} \right)^2} \left[ -\pi + \frac{\tau^2 - 1}{\tau} l\left(\frac{R}{r_\mathrm{s}}; \tau\right) \right] \right\}, & \tau \geq 1
        \\
        \frac{\tau^4}{(2\tau^2 + 1)^3} \left\{ 2\left[ \tau^2 + 1 + 4\left(\left( \frac{R}{r_\mathrm{s}} \right)^2 - 1\right) \right]  f\left(\frac{R}{r_\mathrm{s}}\right) \right. \\
        \left. \hspace{35pt} +\frac{1}{\tau} \left[ \pi (3\tau^2 - 1) + 2\tau(\tau^2 - 3)\ln\tau \right] \right. \\
        \left. \hspace{35pt} + \frac{1}{\tau^3\sqrt{\tau^2 + \left( \frac{R}{r_\mathrm{s}} \right)^2 }} \left[ -\tau^3\pi \left[ 4\left(\tau^2 + \left( \frac{R}{r_\mathrm{s}} \right)^2 \right) - \tau^2 - 1 \right] \right. \right. \\
        \left. \left. \hspace{35pt} \left[ \tau^2(\tau^4 - 1) + \left(\tau^2 + \left( \frac{R}{r_\mathrm{s}} \right)^2 \right)(3\tau^4 - 6\tau^2 - 1) \right] \right] l\left(\frac{R}{r_\mathrm{s}}; \tau\right) \right\}, & \tau < 1
    \end{cases}
\end{align*}
\begin{align*}
    \\
    l\left(\frac{R}{r_\mathrm{s}}; \tau\right) &= \ln\left(\frac{\frac{R}{r_\mathrm{s}}}{\sqrt{\tau^2 + \left(\frac{R}{r_\mathrm{s}}\right)^2} + \tau}\right),
    &
    f\left(\frac{R}{r_\mathrm{s}}\right) &= \begin{cases}
        \frac{-\arccos\left(\frac{R}{r_\mathrm{s}}\right)}{\sqrt{\left(\frac{R}{r_\mathrm{s}}\right)^2 - 1}} & \text{if } R < r_\mathrm{s}
        \\
        1 & \text{if } R = r_\mathrm{s}
        \\
        \frac{\arccos\left(\frac{R}{r_\mathrm{s}}\right)}{\sqrt{\left(\frac{R}{r_\mathrm{s}}\right)^2 - 1}} & \text{if } R > r_\mathrm{s}
    \end{cases}
\end{align*}
and $\delta_\mathrm{c}(c)$ as defined in Eq.~\ref{eq:delta_c}.
This function has three free parameters, which can be chosen to be $c$, $\tau = r_\mathrm{t}/r_\mathrm{s}$, and $r_\mathrm{s} = r_{200}/c$.
The following steps were taken to find the best-fit tNFW parameters:

\begin{enumerate}
    \item Calculate $M_\mathrm{cut} = M_\mathrm{dPIE}(r_\mathrm{cut})$.
    We constrain the tNFW fit to conserve this quantity, since this is the mass constraint obtained from our lensing mass models.
    
    \item $M_\mathrm{cut}$ thus constrained, we recast the 2D tNFW mass profile as a function of two free parameters: $r_\mathrm{s} = r_{200}/c$ and $\tau = r_\mathrm{t}/r_\mathrm{s}$: \begin{equation*}
        M_\mathrm{2D}(R) = 4\pi r_\mathrm{s}^3 \rho_\mathrm{c} \times \delta_\mathrm{c}(r_\mathrm{s}; \tau, r_\mathrm{cut}, M_\mathrm{cut}) \times \beta\left(\frac{R}{r_\mathrm{s}};\tau\right)
    \end{equation*} with \begin{align*}
        \delta_\mathrm{c}(r_\mathrm{s}; \tau, r_\mathrm{cut}, M_\mathrm{cut}) &= \frac{M_\mathrm{cut}}{4\pi r_\mathrm{s}^3 \rho_\mathrm{c}\beta\left(\frac{r_\mathrm{cut}}{r_\mathrm{s}} ; \tau\right)}
    \end{align*}
    
    \item Three points $\log R_i$ are randomly selected from a log-scale uniform distribution over the interval $\left(\log r_0 + [0.2 \times (\log r_\mathrm{cut} - \log r_0)], \log r_\mathrm{cut}\right)$.
    The points are constrained to not be within $0.075 \times \left(\log r_\mathrm{cut} - \log r_0\right)$ of each other.
    The points $R_i$ will be used for a least-squared fitting procedure.
    See Fig.~\ref{fig:example_fit}.
    Note that with two data points, together with the mass conservation constraint at $r_\mathrm{cut}$, a tNFW profile can be fit exactly, since it has three free parameters.
    Three data points is found to be optimal, since it gives significant improvements on the error from two data points, while four or more data points do not significantly tighten the error bars, yet increase the computation time.
    
    \item We build an initial parameter space in which $r_\mathrm{s}\in\left(0.01 \text{ kpc}, r_\mathrm{cut}\right)$ and $\tau\in\left(0.01, 1000\right)$.
    Note that the parameters may settle beyond these initial intervals.
    Only the conditions $r_\mathrm{s} > 0$ and $\tau > 0$ are enforced.
    The parameter space is a $100\times100$ grid.

    \item For each cell in the parameter space, the squared error is calculated as \begin{equation*}
        \Sigma_i \left( \log M_\mathrm{2D}(R_i) - \log M_\mathrm{dPIE}(R_i) \right)^2.
    \end{equation*} The cell with the least squared error is chosen as the center of the zoomed in parameter space.

    \item A zoomed in $100\times100$ parameter space is constructed by setting the limits to be $2\times$ the distance from the last parameter space center.
    For the first iteration, the limits are chosen to be $\pm10\%$ of the best-fit values.

    \item This process is repeated until subsequent best-fit values agree to within 0.01\% (i.e., they agree to four significant figures: $|\tau_\mathrm{new} - \tau_\mathrm{old}| < \tau_\mathrm{old}\times10^{-4}$ and $|r_\mathrm{s,new} - r_\mathrm{s,old}| < r_\mathrm{s,old}\times10^{-4}$).

    \item Once a best-fit $r_\mathrm{s}$ and $\tau$ are found, the best-fit concentration parameter is numerically derived from $\delta_\mathrm{c}(r_\mathrm{s}; \tau, r_\mathrm{cut, M_\mathrm{cut}})$ using a bisection method.

    \item To generate confidence intervals, this process is repeated for each potential 500 times, with different values of $r_i$ randomly chosen, as described in step 3, at each iteration.
    This gives us the range of $c$, $r_\mathrm{s}$, and $\tau$ values that fit well to the \texttt{LENSTOOL} dPIE fits while also conserving mass within $r_\mathrm{cut}$.
\end{enumerate}

\section{Relationship between GGSL and Einstein radius}
\label{sec:app_einstein_radius}
\begin{figure}
    \centering
    \includegraphics[width=1.0\textwidth]{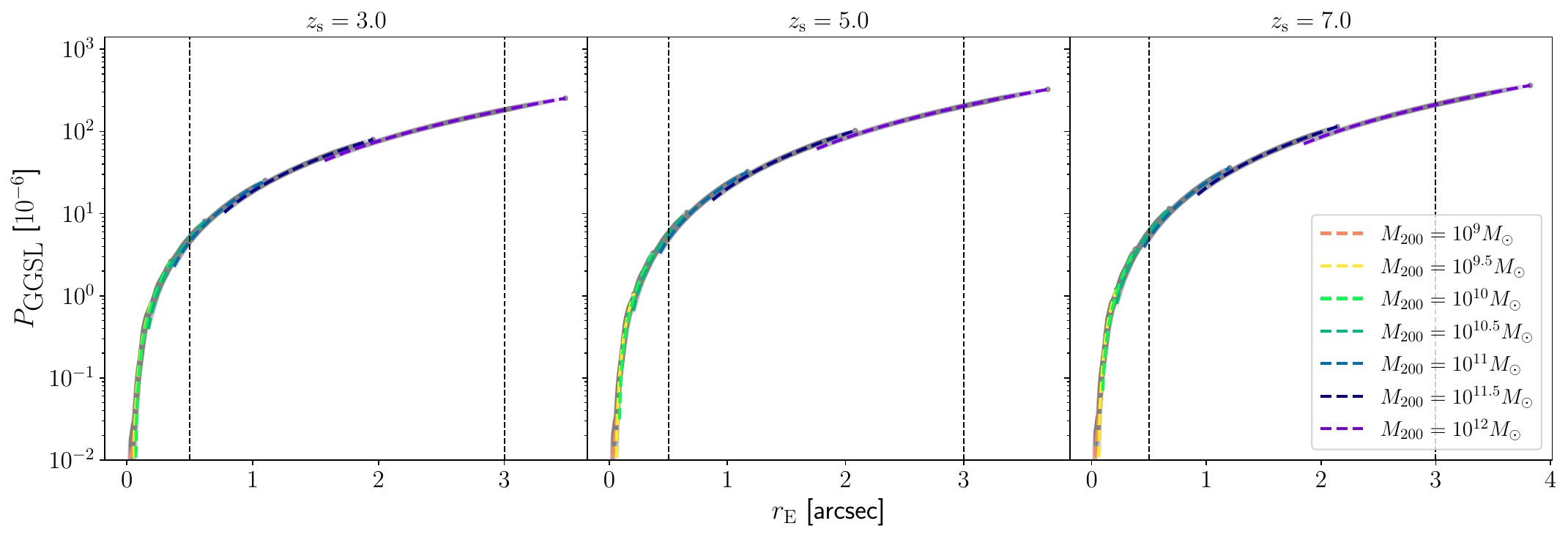}
    \caption{The relationship between the Einstein radius of an isolated sub-halo against the large scale background of AS1063 and the resulting GGSL probability for $z_\mathrm{s}\in\{3,5,7\}$.  We look at five sub-halos in AS1063 representing five mass bins in order to cover the full parameter space. We change the $r_\mathrm{E}$ while keeping the mass of the sub-halo constant by adjusting the inner and outer slopes of a gNFW profile.  Each dashed curve represents the best-fit third degree polynomial to the data for one $M_\mathrm{vir}$, with the data points shown in a light grey scatter behind it.  The data follow each curve very tightly.  The black dashed lines demarcate the $r_\mathrm{E}$ range that we have adopted to define secondary critical lines (Sec.~\ref{subsec:ggsl_def}).} 
    \label{fig:r_e_vs_ggsl}
\end{figure}
Since GGSL cross-section (and in turn, GGSL probability) is directly proportional to the total area enclosed by secondary caustics in the source plane, we expect the Einstein radius, which measures the area enclosed within the critical line due of an isolated sub-halo, to increase and decrease together with the GGSL probability. 
In order to demonstrate and quantify this correlation, we analyze a single simulated sub-halo against the large scale background of the cluster AS1063. 
In this simple construction, the sub-halo gives rise to a single secondary critical line, while the primary critical line is due to the cluster scale halos. 
For a given sub-halo mass, we model the sub-halo using a generalized NFW (gNFW) profile \citep{munozCuspedMassModels2001}.
By varying the gNFW inner and outer slope, we change the Einstein radius of the sub-halo while conserving the sub-halo's virial mass, and track the corresponding change in the GGSL probability (note that both the GGSL probability and FoV caustic change as the sub-halo profile is varied).
Note that these gNFW profiles are not necessarily valid within a \lcdm\ framework, but are used here to illustrate the relationship between $r_\mathrm{E}$ and $P_\mathrm{GGSL}$. 
We repeat this for $\log\left(M_\mathrm{200}/M_\odot\right)\in \{9, 9.5, 10, 10.5, 11, 11.5, 12\}$, and for source plane redshifts $z_\mathrm{s}\in\{3,5,7\}$.

Results are shown in Fig.~\ref{fig:r_e_vs_ggsl}.
The Einstein radius and GGSL probability are seen to correlate tightly at all scales. We found third degree polynomial fits for each mass bin.
In the secondary critical line band from $r_\mathrm{E} = 0.5''$ to $r_\mathrm{E} = 3.0''$, there is a power law-like relation.


\bibliography{ggsl_bib_zotero}{}
\bibliographystyle{aasjournal}

\end{document}

%% file: authors.tex
\correspondingauthor{Yarone M. Tokayer}
\email{yarone.tokayer@yale.edu}

\author[0000-0002-0430-5798]{Yarone M. Tokayer}
\affiliation{Department of Physics, Yale University, P.O. Box 208120, New Haven, CT 06520, USA}

\author[0000-0001-7040-4930]{Isaque Dutra}
\affiliation{Department of Physics, Yale University, P.O. Box 208120, New Haven, CT 06520, USA}

\author[0000-0002-5554-8896]{Priyamvada Natarajan}
\affiliation{Department of Astronomy, Yale University, P.O. Box 208101,
New Haven, CT 06520, USA}
\affiliation{Department of Physics, Yale University, P.O. Box 208120, New Haven, CT 06520, USA}
\affiliation{Black Hole Initiative, 20 Garden Street, 2nd Floor, Cambridge, MA 02138, USA}

\author[0000-0003-3266-2001]{Guillaume Mahler}
\affiliation{Centre for Extragalactic Astronomy, Durham University, South Road, Durham DH1 3LE, UK}
\affiliation{Institute for Computational Cosmology, Durham University, South Road, Durham DH1 3LE, UK}

\author[0000-0003-1974-8732]{Mathilde Jauzac}
\affiliation{Centre for Extragalactic Astronomy, Durham University, South Road, Durham DH1 3LE, UK}
\affiliation{Institute for Computational Cosmology, Durham University, South Road, Durham DH1 3LE, UK}
\affiliation{Astrophysics Research Centre, University of KwaZulu-Natal, Westville Campus, Durban 4041, South Africa}
\affiliation{School of Mathematics, Statistics \& Computer Science, University of KwaZulu-Natal, Westville Campus, Durban 4041, South Africa}

\author[0000-0003-1225-7084]{Massimo Meneghetti}
\affiliation{INAF-OAS, Osservatorio di Astrofisica e Scienza dello Spazio di Bologna, via Gobetti 93/3, I-40129 Bologna, Italy}
\affiliation{National Institute for Nuclear Physics, viale Berti Pichat 6/2, I-40127 Bologna, Italy}

%% file: abstract.tex
\begin{abstract}
Strong gravitational lensing offers a powerful probe of the detailed distribution of matter in lenses, while magnifying and bringing faint background sources into view.
Observed strong lensing by massive galaxy clusters, which are often in complex dynamical states, has also been used to map their dark matter substructures on smaller scales. 
Deep high resolution imaging has revealed the presence of strong lensing events associated with these substructures, namely galaxy-scale sub-halos.
However, an inventory of these observed galaxy-galaxy strong lensing (GGSL) events is noted to be discrepant with state-of-the-art \lcdm\ simulations.
Cluster sub-halos appear to be over-concentrated compared to their simulated counterparts yielding an order of magnitude higher value of GGSL.
In this paper, we explore the possibility of resolving this observed discrepancy by redistributing the mass within observed cluster sub-halos in ways that are consistent within the \lcdm\ paradigm of structure formation.
Lensing mass reconstructions from data provide constraints on the mass enclosed within apertures and are agnostic to the detailed mass profile within them.
Therefore, as the detailed density profile within cluster sub-halos currently remains unconstrained by data, we are afforded the freedom to redistribute the enclosed mass. 
We investigate if rearranging the mass to a more centrally concentrated density profile helps alleviate the GGSL discrepancy.
We report that refitting cluster sub-halos to the ubiquitous \lcdm-motivated Navarro-Frenk-White profile, and further modifying them to include significant baryonic components, does not resolve this tension.
A resolution to this persisting GGSL discrepancy may require more careful exploration of alternative dark matter models.

\end{abstract}

%% file: cluster_samples_table.tex
\begin{deluxetable}{lcCCc}
    \tablecaption{Summary of cluster lenses
    \label{tab:cluster_info}}
    \tablehead{ \colhead{Cluster name} & \colhead{Redshift} & \colhead{Reported mass estimate} & \mathrm{Mass}\ ($10^{15}\ M_\odot\ h^{-1}$)$^a$ & \colhead{Galaxy-scale sub-halos} 
    }
    \colnumbers
    \startdata
        Abell S1063 & 0.348 & M_{200} & 2.90\pm1.33 & 222 
        \\
        MACS J0416.1-2403 & 0.397 & M_\mathrm{vir}\ ^b & 0.909 \pm 0.230 & 191 
        \\
        MACS J1206.2-0847 & 0.440 & M_\mathrm{vir}\ ^c & 1.1 \pm 0.3 & 258 
        \\
        Abell 2744 & 0.308 & M_\mathrm{enc}(\text{0.9 Mpc}\ h^{-1}) & 1.6\pm0.1 & 246 
        \\
        PSZ1 G311.65-18.48 & 0.443 & M_\mathrm{enc}(\text{175 kpc}\ h^{-1}) & 0.205^{+0.001}_{-0.001} & 194 
    \enddata

    \tablecomments{Column 5 includes sub-halos that are optimized using the dPIE scaling relations used for that cluster.
    \\
    $^a$See text for references to each cluster. $^b$ See table 3 of \cite{umetsuCLASHJOINTANALYSIS2016}.  $^c$ $\Delta_\mathrm{vir}\approx 132$. 
    }
\end{deluxetable}

%% file: main.bbl
\begin{thebibliography}{}
\expandafter\ifx\csname natexlab\endcsname\relax\def\natexlab#1{#1}\fi
\providecommand{\url}[1]{\href{#1}{#1}}
\providecommand{\dodoi}[1]{doi:~\href{http://doi.org/#1}{\nolinkurl{#1}}}
\providecommand{\doeprint}[1]{\href{http://ascl.net/#1}{\nolinkurl{http://ascl.net/#1}}}
\providecommand{\doarXiv}[1]{\href{https://arxiv.org/abs/#1}{\nolinkurl{https://arxiv.org/abs/#1}}}

\bibitem[{Abell {et~al.}(1989)Abell, Corwin, \&
  Olowin}]{abellCatalogRichClusters1989}
Abell, G.~O., Corwin, Jr., H.~G., \& Olowin, R.~P. 1989, The Astrophysical
  Journal Supplement Series, 70, 1, \dodoi{10.1086/191333}

\bibitem[{Amruth {et~al.}(2023)Amruth, Broadhurst, Lim, Oguri, Smoot, Diego,
  Leung, Emami, Li, Chiueh, Schive, Yeung, \&
  Li}]{amruthEinsteinRingsModulated2023}
Amruth, A., Broadhurst, T., Lim, J., {et~al.} 2023, Nat Astron, 7, 736,
  \dodoi{10.1038/s41550-023-01943-9}

\bibitem[{Andrade {et~al.}(2023)Andrade, Kaplinghat, \&
  Valli}]{andradeHaloDensitiesPericenter2023}
Andrade, K.~E., Kaplinghat, M., \& Valli, M. 2023, Halo {{Densities}} and
  {{Pericenter Distances}} of the {{Bright Milky Way Satellites}} as a {{Test}}
  of {{Dark Matter Physics}},  arXiv, \dodoi{10.48550/arXiv.2311.01528}

\bibitem[{{Astropy Collaboration} {et~al.}(2013){Astropy Collaboration},
  Robitaille, Tollerud, Greenfield, Droettboom, Bray, Aldcroft, Davis,
  Ginsburg, {Price-Whelan}, Kerzendorf, Conley, Crighton, Barbary, Muna,
  Ferguson, Grollier, Parikh, Nair, Unther, Deil, Woillez, Conseil, Kramer,
  Turner, Singer, Fox, Weaver, Zabalza, Edwards, Azalee~Bostroem, Burke, Casey,
  Crawford, Dencheva, Ely, Jenness, Labrie, Lim, Pierfederici, Pontzen, Ptak,
  Refsdal, Servillat, \&
  Streicher}]{astropycollaborationAstropyCommunityPython2013}
{Astropy Collaboration}, Robitaille, T.~P., Tollerud, E.~J., {et~al.} 2013,
  {\textbackslash}aap, 558, A33, \dodoi{10.1051/0004-6361/201322068}

\bibitem[{{Astropy Collaboration} {et~al.}(2018){Astropy Collaboration},
  {Price-Whelan}, Sip{\textbackslash}Hocz, G{\"u}nther, Lim, Crawford, Conseil,
  Shupe, Craig, Dencheva, Ginsburg, {Vand erPlas}, Bradley,
  {P{\'e}rez-Su{\'a}rez}, {de Val-Borro}, Aldcroft, Cruz, Robitaille, Tollerud,
  Ardelean, Babej, Bach, Bachetti, Bakanov, Bamford, Barentsen, Barmby,
  Baumbach, Berry, Biscani, Boquien, Bostroem, Bouma, Brammer, Bray,
  Breytenbach, Buddelmeijer, Burke, Calderone, Cano~Rodr{\'i}guez, Cara,
  Cardoso, Cheedella, Copin, Corrales, Crichton, D'Avella, Deil, Depagne,
  Dietrich, Donath, Droettboom, Earl, Erben, Fabbro, Ferreira, Finethy, Fox,
  Garrison, Gibbons, Goldstein, Gommers, Greco, Greenfield, Groener, Grollier,
  Hagen, Hirst, Homeier, Horton, Hosseinzadeh, Hu, Hunkeler, Ivezi{\'c}, Jain,
  Jenness, Kanarek, Kendrew, Kern, Kerzendorf, Khvalko, King, Kirkby, Kulkarni,
  Kumar, Lee, Lenz, Littlefair, Ma, Macleod, Mastropietro, McCully, Montagnac,
  Morris, Mueller, Mumford, Muna, Murphy, Nelson, Nguyen, Ninan, N{\"o}the,
  Ogaz, Oh, Parejko, Parley, Pascual, Patil, Patil, Plunkett, Prochaska,
  Rastogi, Reddy~Janga, Sabater, Sakurikar, Seifert, Sherbert,
  {Sherwood-Taylor}, Shih, Sick, Silbiger, Singanamalla, Singer, Sladen,
  Sooley, Sornarajah, Streicher, Teuben, Thomas, Tremblay, Turner, Terr{\'o}n,
  {van Kerkwijk}, {de la Vega}, Watkins, Weaver, Whitmore, Woillez, Zabalza, \&
  {Astropy Contributors}}]{astropycollaborationAstropyProjectBuilding2018}
{Astropy Collaboration}, {Price-Whelan}, A.~M., Sip{\textbackslash}Hocz, B.~M.,
  {et~al.} 2018, {\textbackslash}aj, 156, 123, \dodoi{10.3847/1538-3881/aabc4f}

\bibitem[{{Astropy Collaboration} {et~al.}(2022){Astropy Collaboration},
  {Price-Whelan}, Lim, Earl, Starkman, Bradley, Shupe, Patil, Corrales,
  Brasseur, N{\"o}the, Donath, Tollerud, Morris, Ginsburg, Vaher, Weaver,
  Tocknell, Jamieson, {van Kerkwijk}, Robitaille, Merry, Bachetti, G{\"u}nther,
  Aldcroft, {Alvarado-Montes}, Archibald, B{\'o}di, Bapat, Barentsen,
  Baz{\'a}n, Biswas, Boquien, Burke, Cara, Cara, Conroy, Conseil, Craig, Cross,
  Cruz, D'Eugenio, Dencheva, Devillepoix, Dietrich, Eigenbrot, Erben, Ferreira,
  {Foreman-Mackey}, Fox, Freij, Garg, Geda, Glattly, Gondhalekar, Gordon,
  Grant, Greenfield, Groener, Guest, Gurovich, Handberg, Hart,
  {Hatfield-Dodds}, Homeier, Hosseinzadeh, Jenness, Jones, Joseph, Kalmbach,
  Karamehmetoglu, Ka{\l}uszy{\'n}ski, Kelley, Kern, Kerzendorf, Koch, Kulumani,
  Lee, Ly, Ma, MacBride, Maljaars, Muna, Murphy, Norman, O'Steen, Oman,
  Pacifici, Pascual, {Pascual-Granado}, Patil, Perren, Pickering, Rastogi,
  Roulston, Ryan, Rykoff, Sabater, Sakurikar, Salgado, Sanghi, Saunders,
  Savchenko, Schwardt, {Seifert-Eckert}, Shih, Jain, Shukla, Sick, Simpson,
  Singanamalla, Singer, Singhal, Sinha, Sip{\H o}cz, Spitler, Stansby,
  Streicher, {\v S}umak, Swinbank, Taranu, Tewary, Tremblay, {de Val-Borro},
  Van~Kooten, Vasovi{\'c}, Verma, {de Miranda Cardoso}, Williams, Wilson,
  Winkel, {Wood-Vasey}, Xue, Yoachim, Zhang, Zonca, \& {Astropy Project
  Contributors}}]{astropycollaborationAstropyProjectSustaining2022}
{Astropy Collaboration}, {Price-Whelan}, A.~M., Lim, P.~L., {et~al.} 2022, The
  Astrophysical Journal, 935, 167, \dodoi{10.3847/1538-4357/ac7c74}

\bibitem[{Balestra {et~al.}(2016)Balestra, Mercurio, Sartoris, Girardi, Grillo,
  Nonino, Rosati, Biviano, Ettori, Forman, Jones, Koekemoer, Medezinski,
  Merten, Ogrean, Tozzi, Umetsu, Vanzella, van Weeren, Zitrin, Annunziatella,
  Caminha, Broadhurst, Coe, Donahue, Fritz, Frye, Kelson, Lombardi, Maier,
  Meneghetti, Monna, Postman, Scodeggio, Seitz, \&
  Ziegler}]{balestraCLASHVLTDISSECTINGFRONTIER2016}
Balestra, I., Mercurio, A., Sartoris, B., {et~al.} 2016, ApJS, 224, 33,
  \dodoi{10.3847/0067-0049/224/2/33}

\bibitem[{Baltz {et~al.}(2009)Baltz, Marshall, \&
  Oguri}]{baltzAnalyticModelsPlausible2009}
Baltz, E.~A., Marshall, P., \& Oguri, M. 2009, J. Cosmol. Astropart. Phys.,
  2009, 015, \dodoi{10.1088/1475-7516/2009/01/015}

\bibitem[{Banik \& van~den Bosch(2022)}]{banikDynamicalFrictionBuoyancy2022}
Banik, U., \& van~den Bosch, F.~C. 2022, ApJ, 926, 215,
  \dodoi{10.3847/1538-4357/ac4242}

\bibitem[{Beauchesne {et~al.}(2023)Beauchesne, Cl{\'e}ment, Hibon, Limousin,
  Eckert, Kneib, Richard, Natarajan, Jauzac, Montes, Mahler, Claeyssens,
  Jeanneau, Koekemoer, Lagattuta, Pagul, \&
  S{\'a}nchez}]{beauchesneNewStepForward2023}
Beauchesne, B., Cl{\'e}ment, B., Hibon, P., {et~al.} 2023, A New Step Forward
  in Realistic Cluster Lens Mass Modelling: {{Analysis}} of {{Hubble Frontier
  Field Cluster Abell S1063}} from Joint Lensing, {{X-ray}} and Galaxy
  Kinematics Data, \dodoi{10.48550/arXiv.2301.10907}

\bibitem[{Beck {et~al.}(2016)Beck, Murante, Arth, Remus, Teklu, Donnert,
  Planelles, Beck, F{\"o}rster, Imgrund, Dolag, \&
  Borgani}]{beckImprovedSPHScheme2016}
Beck, A.~M., Murante, G., Arth, A., {et~al.} 2016, Monthly Notices of the Royal
  Astronomical Society, 455, 2110, \dodoi{10.1093/mnras/stv2443}

\bibitem[{Bergamini {et~al.}(2019)Bergamini, Rosati, Mercurio, Grillo, Caminha,
  Meneghetti, Agnello, Biviano, Calura, Giocoli, Lombardi, Rodighiero, \&
  Vanzella}]{bergaminiEnhancedClusterLensing2019}
Bergamini, P., Rosati, P., Mercurio, A., {et~al.} 2019, A\&A, 631, A130,
  \dodoi{10.1051/0004-6361/201935974}

\bibitem[{Bergamini {et~al.}(2021)Bergamini, Rosati, Vanzella, Caminha, Grillo,
  Mercurio, Meneghetti, Angora, Calura, Nonino, \&
  Tozzi}]{bergaminiNewHighprecisionStrong2021}
Bergamini, P., Rosati, P., Vanzella, E., {et~al.} 2021, A\&A, 645, A140,
  \dodoi{10.1051/0004-6361/202039564}

\bibitem[{Bezanson {et~al.}(2022)Bezanson, Labbe, Whitaker, Leja, Price, Franx,
  Brammer, Marchesini, Zitrin, Wang, Weaver, Furtak, Atek, Coe, Cutler, Dayal,
  {van Dokkum}, Feldmann, Forster~Schreiber, Fujimoto, Geha, Glazebrook, {de
  Graaff}, Greene, Juneau, Kassin, Kriek, Khullar, Maseda, Mowla, Muzzin,
  Nanayakkara, Nelson, Oesch, Pacifici, Pan, Papovich, Setton, Shapley, Smit,
  Stefanon, Taylor, \& Williams}]{bezansonJWSTUNCOVERTreasury2022}
Bezanson, R., Labbe, I., Whitaker, K.~E., {et~al.} 2022, The {{JWST UNCOVER
  Treasury}} Survey: {{Ultradeep NIRSpec}} and {{NIRCam ObserVations}} before
  the {{Epoch}} of {{Reionization}}, \dodoi{10.48550/arXiv.2212.04026}

\bibitem[{Birrer \&
  Amara(2018)}]{birrerLenstronomyMultipurposeGravitational2018}
Birrer, S., \& Amara, A. 2018, Astrophysics Source Code Library, ascl:1804.012

\bibitem[{Birrer {et~al.}(2021)Birrer, Shajib, Gilman, Galan, Aalbers, Millon,
  Morgan, Pagano, Park, Teodori, Tessore, Ueland, de~Vyvere, {Wagner-Carena},
  Wempe, Yang, Ding, Schmidt, Sluse, Zhang, \&
  Amara}]{birrerLenstronomyIIGravitational2021}
Birrer, S., Shajib, A.~J., Gilman, D., {et~al.} 2021, Journal of Open Source
  Software, 6, 3283, \dodoi{10.21105/joss.03283}

\bibitem[{Bogd{\'a}n {et~al.}(2023)Bogd{\'a}n, Goulding, Natarajan, Kov{\'a}cs,
  Tremblay, Chadayammuri, Volonteri, Kraft, Forman, Jones, Churazov, \&
  Zhuravleva}]{bogdanEvidenceHeavyseedOrigin2023}
Bogd{\'a}n, {\'A}., Goulding, A.~D., Natarajan, P., {et~al.} 2023, Nat Astron,
  1, \dodoi{10.1038/s41550-023-02111-9}

\bibitem[{{Boylan-Kolchin} {et~al.}(2011){Boylan-Kolchin}, Bullock, \&
  Kaplinghat}]{boylan-kolchinTooBigFail2011}
{Boylan-Kolchin}, M., Bullock, J.~S., \& Kaplinghat, M. 2011, Monthly Notices
  of the Royal Astronomical Society, 415, L40,
  \dodoi{10.1111/j.1745-3933.2011.01074.x}

\bibitem[{Bullock \&
  {Boylan-Kolchin}(2017)}]{bullockSmallScaleChallengesLCDM2017}
Bullock, J.~S., \& {Boylan-Kolchin}, M. 2017, Annual Review of Astronomy and
  Astrophysics, 55, 343, \dodoi{10.1146/annurev-astro-091916-055313}

\bibitem[{Bullock {et~al.}(2001)Bullock, Kolatt, Sigad, Somerville, Kravtsov,
  Klypin, Primack, \& Dekel}]{bullockProfilesDarkHaloes2001}
Bullock, J.~S., Kolatt, T.~S., Sigad, Y., {et~al.} 2001, Monthly Notices of the
  Royal Astronomical Society, 321, 559,
  \dodoi{10.1046/j.1365-8711.2001.04068.x}

\bibitem[{Cui {et~al.}(2018)Cui, Knebe, Yepes, Pearce, Power, Dave, Arth,
  Borgani, Dolag, Elahi, Mostoghiu, Murante, Rasia, Stoppacher, {Vega-Ferrero},
  Wang, Yang, Benson, Cora, Croton, Sinha, Stevens, {Vega-Mart{\'i}nez},
  Arthur, Baldi, Ca{\~n}as, Cialone, Cunnama, De~Petris, Durando, Ettori,
  Gottl{\"o}ber, Nuza, Old, Pilipenko, Sorce, \&
  Welker}]{cuiThreeHundredProject2018}
Cui, W., Knebe, A., Yepes, G., {et~al.} 2018, Monthly Notices of the Royal
  Astronomical Society, 480, 2898, \dodoi{10.1093/mnras/sty2111}

\bibitem[{Dalal \& Kochanek(2002)}]{dalalDirectDetectionCold2002}
Dalal, N., \& Kochanek, C.~S. 2002, ApJ, 572, 25, \dodoi{10.1086/340303}

\bibitem[{{de Blok}(2010)}]{deblokCoreCuspProblem2010}
{de Blok}, W. J.~G. 2010, Advances in Astronomy, 2010, 789293,
  \dodoi{10.1155/2010/789293}

\bibitem[{Del~Popolo \& Le~Delliou(2017)}]{delpopoloSmallScaleProblems2017}
Del~Popolo, A., \& Le~Delliou, M. 2017, Galaxies, 5, 17,
  \dodoi{10.3390/galaxies5010017}

\bibitem[{Diemer \& Kravtsov(2015)}]{diemerUniversalModelHalo2015}
Diemer, B., \& Kravtsov, A.~V. 2015, The Astrophysical Journal, 799, 108,
  \dodoi{10.1088/0004-637X/799/1/108}

\bibitem[{Ebeling {et~al.}(2001)Ebeling, Edge, \&
  Henry}]{ebelingMACSQuestMost2001}
Ebeling, H., Edge, A.~C., \& Henry, J.~P. 2001, ApJ, 553, 668,
  \dodoi{10.1086/320958}

\bibitem[{Eichner {et~al.}(2013)Eichner, Seitz, Suyu, Halkola, Umetsu, Zitrin,
  Coe, Monna, Rosati, Grillo, Balestra, Postman, Koekemoer, Zheng, H{\o}st,
  Lemze, Broadhurst, Moustakas, Bradley, Molino, Nonino, Mercurio, Scodeggio,
  Bartelmann, Benitez, Bouwens, Donahue, Infante, Jouvel, Kelson, Lahav,
  Medezinski, Melchior, Merten, \& Riess}]{eichnerGALAXYHALOTRUNCATION2013}
Eichner, T., Seitz, S., Suyu, S.~H., {et~al.} 2013, ApJ, 774, 124,
  \dodoi{10.1088/0004-637X/774/2/124}

\bibitem[{Eke {et~al.}(2001)Eke, Navarro, \&
  Steinmetz}]{ekePowerSpectrumDependence2001}
Eke, V.~R., Navarro, J.~F., \& Steinmetz, M. 2001, ApJ, 554, 114,
  \dodoi{10.1086/321345}

\bibitem[{El{\'i}asd{\'o}ttir {et~al.}(2007)El{\'i}asd{\'o}ttir, Limousin,
  Richard, Hjorth, Kneib, Natarajan, Pedersen, Jullo, \&
  Paraficz}]{eliasdottirWhereMatterMerging2007}
El{\'i}asd{\'o}ttir, {\'A}., Limousin, M., Richard, J., {et~al.} 2007, Where Is
  the Matter in the {{Merging Cluster Abell}} 2218?,
  \dodoi{10.48550/arXiv.0710.5636}

\bibitem[{Gilman {et~al.}(2020{\natexlab{a}})Gilman, Birrer, Nierenberg, Treu,
  Du, \& Benson}]{gilmanWarmDarkMatter2020}
Gilman, D., Birrer, S., Nierenberg, A., {et~al.} 2020{\natexlab{a}}, Monthly
  Notices of the Royal Astronomical Society, 491, 6077,
  \dodoi{10.1093/mnras/stz3480}

\bibitem[{Gilman {et~al.}(2020{\natexlab{b}})Gilman, Du, Benson, Birrer,
  Nierenberg, \& Treu}]{gilmanConstraintsMassConcentration2020}
Gilman, D., Du, X., Benson, A., {et~al.} 2020{\natexlab{b}}, Monthly Notices of
  the Royal Astronomical Society: Letters, 492, L12,
  \dodoi{10.1093/mnrasl/slz173}

\bibitem[{Giocoli {et~al.}(2012)Giocoli, Meneghetti, Bartelmann, Moscardini, \&
  Boldrin}]{giocoliMokaNewTool2012}
Giocoli, C., Meneghetti, M., Bartelmann, M., Moscardini, L., \& Boldrin, M.
  2012, Monthly Notices of the Royal Astronomical Society, 421, 3343,
  \dodoi{10.1111/j.1365-2966.2012.20558.x}

\bibitem[{Girelli {et~al.}(2020)Girelli, Pozzetti, Bolzonella, Giocoli,
  Marulli, \& Baldi}]{girelliStellartohaloMassRelation2020}
Girelli, G., Pozzetti, L., Bolzonella, M., {et~al.} 2020, A\&A, 634, A135,
  \dodoi{10.1051/0004-6361/201936329}

\bibitem[{Gnedin {et~al.}(2004)Gnedin, Kravtsov, Klypin, \&
  Nagai}]{gnedinResponseDarkMatter2004}
Gnedin, O.~Y., Kravtsov, A.~V., Klypin, A.~A., \& Nagai, D. 2004, ApJ, 616, 16,
  \dodoi{10.1086/424914}

\bibitem[{G{\'o}mez {et~al.}(2012)G{\'o}mez, Valkonen, Romer, {Lloyd-Davies},
  Verdugo, Cantalupo, Daub, Goldstein, Kuo, Lange, Lueker, Holzapfel, Peterson,
  Ruhl, Runyan, Reichardt, \& Sabirli}]{gomezOPTICALXRAYOBSERVATIONS2012}
G{\'o}mez, P.~L., Valkonen, L.~E., Romer, A.~K., {et~al.} 2012, AJ, 144, 79,
  \dodoi{10.1088/0004-6256/144/3/79}

\bibitem[{Gonzalez {et~al.}(2020)Gonzalez, Chalela, Jauzac, Eckert, Schaller,
  Harvey, Niemiec, Koekemoer, Barnes, Clowe, Connor, Diego, Remolina~Gonzalez,
  \& Steinhardt}]{gonzalezSettingSceneBUFFALO2020}
Gonzalez, E.~J., Chalela, M., Jauzac, M., {et~al.} 2020, Monthly Notices of the
  Royal Astronomical Society, 494, 349, \dodoi{10.1093/mnras/staa745}

\bibitem[{Gralla {et~al.}(2011)Gralla, Sharon, Gladders, Marrone, Barrientos,
  Bayliss, Bonamente, Bulbul, Carlstrom, Culverhouse, Gilbank, Greer, Hasler,
  Hawkins, Hennessy, Joy, Koester, Lamb, Leitch, Miller, Mroczkowski, Muchovej,
  Oguri, Plagge, Pryke, \& Woody}]{grallaSunyaevZelDovichEffect2011}
Gralla, M.~B., Sharon, K., Gladders, M.~D., {et~al.} 2011, The Astrophysical
  Journal, 737, 74, \dodoi{10.1088/0004-637X/737/2/74}

\bibitem[{Green \& {van~den~Bosch}(2019)}]{greenTidalEvolutionDark2019}
Green, S.~B., \& {van~den~Bosch}, F.~C. 2019, Monthly Notices of the Royal
  Astronomical Society, 490, 2091, \dodoi{10.1093/mnras/stz2767}

\bibitem[{Green {et~al.}(2021)Green, {van~den~Bosch}, \&
  Jiang}]{greenTidalEvolutionDark2021}
Green, S.~B., {van~den~Bosch}, F.~C., \& Jiang, F. 2021, Monthly Notices of the
  Royal Astronomical Society, 503, 4075, \dodoi{10.1093/mnras/stab696}

\bibitem[{Gruen {et~al.}(2013)Gruen, Brimioulle, Seitz, Lee, Young,
  Koppenhoefer, Eichner, Riffeser, Vikram, Weidinger, \&
  Zenteno}]{gruenWeakLensingAnalysis2013}
Gruen, D., Brimioulle, F., Seitz, S., {et~al.} 2013, Monthly Notices of the
  Royal Astronomical Society, 432, 1455, \dodoi{10.1093/mnras/stt566}

\bibitem[{Hezaveh {et~al.}(2016)Hezaveh, Dalal, Marrone, Mao, Morningstar, Wen,
  Blandford, Carlstrom, Fassnacht, Holder, Kemball, Marshall, Murray,
  Levasseur, Vieira, \& Wechsler}]{hezavehDETECTIONLENSINGSUBSTRUCTURE2016}
Hezaveh, Y.~D., Dalal, N., Marrone, D.~P., {et~al.} 2016, ApJ, 823, 37,
  \dodoi{10.3847/0004-637X/823/1/37}

\bibitem[{Hinshaw {et~al.}(2013)Hinshaw, Larson, Komatsu, Spergel, Bennett,
  Dunkley, Nolta, Halpern, Hill, Odegard, Page, Smith, Weiland, Gold, Jarosik,
  Kogut, Limon, Meyer, Tucker, Wollack, \&
  Wright}]{hinshawNINEYEARWILKINSONMICROWAVE2013}
Hinshaw, G., Larson, D., Komatsu, E., {et~al.} 2013, ApJS, 208, 19,
  \dodoi{10.1088/0067-0049/208/2/19}

\bibitem[{Jauzac {et~al.}(2014)Jauzac, Cl{\'e}ment, Limousin, Richard, Jullo,
  Ebeling, Atek, Kneib, Knowles, Natarajan, Eckert, Egami, Massey, \&
  Rexroth}]{jauzacHubbleFrontierFields2014}
Jauzac, M., Cl{\'e}ment, B., Limousin, M., {et~al.} 2014, Monthly Notices of
  the Royal Astronomical Society, 443, 1549, \dodoi{10.1093/mnras/stu1355}

\bibitem[{Jauzac {et~al.}(2015)Jauzac, Richard, Jullo, Cl{\'e}ment, Limousin,
  Kneib, Ebeling, Natarajan, Rodney, Atek, Massey, Eckert, Egami, \&
  Rexroth}]{jauzacHubbleFrontierFields2015}
Jauzac, M., Richard, J., Jullo, E., {et~al.} 2015, Monthly Notices of the Royal
  Astronomical Society, 452, 1437, \dodoi{10.1093/mnras/stv1402}

\bibitem[{Jauzac {et~al.}(2016)Jauzac, Eckert, Schwinn, Harvey, Baugh,
  Robertson, Bose, Massey, Owers, Ebeling, Shan, Jullo, Kneib, Richard, Atek,
  Cl{\'e}ment, Egami, Israel, Knowles, Limousin, Natarajan, Rexroth, Taylor, \&
  Tchernin}]{jauzacExtraordinaryAmountSubstructure2016}
Jauzac, M., Eckert, D., Schwinn, J., {et~al.} 2016, Monthly Notices of the
  Royal Astronomical Society, 463, 3876, \dodoi{10.1093/mnras/stw2251}

\bibitem[{Johnson {et~al.}(2014)Johnson, Sharon, Bayliss, Gladders, Coe, \&
  Ebeling}]{johnsonLensModelsMagnification2014}
Johnson, T.~L., Sharon, K., Bayliss, M.~B., {et~al.} 2014, The Astrophysical
  Journal, 797, 48, \dodoi{10.1088/0004-637X/797/1/48}

\bibitem[{Jullo \& Kneib(2009)}]{julloMultiscaleClusterLens2009}
Jullo, E., \& Kneib, J.~P. 2009, Monthly Notices of the Royal Astronomical
  Society, 395, 1319, \dodoi{10.1111/j.1365-2966.2009.14654.x}

\bibitem[{Jullo {et~al.}(2007)Jullo, Kneib, Limousin, El{\'i}asd{\'o}ttir,
  Marshall, \& Verdugo}]{julloBayesianApproachStrong2007}
Jullo, E., Kneib, J.~P., Limousin, M., {et~al.} 2007, New Journal of Physics,
  9, 447, \dodoi{10.1088/1367-2630/9/12/447}

\bibitem[{Keeton \& Moustakas(2009)}]{keetonNEWCHANNELDETECTING2009}
Keeton, C.~R., \& Moustakas, L.~A. 2009, ApJ, 699, 1720,
  \dodoi{10.1088/0004-637X/699/2/1720}

\bibitem[{Klypin {et~al.}(1999)Klypin, Kravtsov, Valenzuela, \&
  Prada}]{klypinWhereAreMissing1999}
Klypin, A., Kravtsov, A.~V., Valenzuela, O., \& Prada, F. 1999, The
  Astrophysical Journal, 522, 82, \dodoi{10.1086/307643}

\bibitem[{Kneib {et~al.}(1996)Kneib, Ellis, Smail, Couch, \&
  Sharples}]{kneibHubbleSpaceTelescope1996}
Kneib, J.~P., Ellis, R.~S., Smail, I., Couch, W.~J., \& Sharples, R.~M. 1996,
  The Astrophysical Journal, 471, 643, \dodoi{10.1086/177995}

\bibitem[{Kneib \& Natarajan(2011)}]{kneibClusterLenses2011}
Kneib, J.-P., \& Natarajan, P. 2011, Astron Astrophys Rev, 19, 47,
  \dodoi{10.1007/s00159-011-0047-3}

\bibitem[{Limousin {et~al.}(2005)Limousin, Kneib, \&
  Natarajan}]{limousinConstrainingMassDistribution2005}
Limousin, M., Kneib, J.-P., \& Natarajan, P. 2005, Monthly Notices of the Royal
  Astronomical Society, 356, 309, \dodoi{10.1111/j.1365-2966.2004.08449.x}

\bibitem[{Lotz {et~al.}(2017)Lotz, Koekemoer, Coe, Grogin, Capak, Mack,
  Anderson, Avila, Barker, Borncamp, Brammer, Durbin, Gunning, Hilbert,
  Jenkner, Khandrika, Levay, Lucas, MacKenty, Ogaz, Porterfield, Reid,
  Robberto, Royle, Smith, {Storrie-Lombardi}, Sunnquist, Surace, Taylor,
  Williams, Bullock, Dickinson, Finkelstein, Natarajan, Richard, Robertson,
  Tumlinson, Zitrin, Flanagan, Sembach, Soifer, \&
  Mountain}]{lotzFrontierFieldsSurvey2017}
Lotz, J.~M., Koekemoer, A., Coe, D., {et~al.} 2017, ApJ, 837, 97,
  \dodoi{10.3847/1538-4357/837/1/97}

\bibitem[{Ludlow {et~al.}(2016)Ludlow, Bose, Angulo, Wang, Hellwing, Navarro,
  Cole, \& Frenk}]{ludlowMassConcentrationRedshift2016}
Ludlow, A.~D., Bose, S., Angulo, R.~E., {et~al.} 2016, Mon. Not. R. Astron.
  Soc., 460, 1214, \dodoi{10.1093/mnras/stw1046}

\bibitem[{Macci{\`o} {et~al.}(2008)Macci{\`o}, Dutton, \& {van den
  Bosch}}]{maccioConcentrationSpinShape2008}
Macci{\`o}, A.~V., Dutton, A.~A., \& {van den Bosch}, F.~C. 2008, Monthly
  Notices of the Royal Astronomical Society, 391, 1940,
  \dodoi{10.1111/j.1365-2966.2008.14029.x}

\bibitem[{Mahler {et~al.}(2018)Mahler, Richard, Cl{\'e}ment, Lagattuta,
  Schmidt, Patr{\'i}cio, Soucail, Bacon, Pello, Bouwens, Maseda, Martinez,
  Carollo, Inami, Leclercq, \& Wisotzki}]{mahlerStronglensingAnalysisA27442018}
Mahler, G., Richard, J., Cl{\'e}ment, B., {et~al.} 2018, Monthly Notices of the
  Royal Astronomical Society, 473, 663, \dodoi{10.1093/mnras/stx1971}

\bibitem[{Mantz {et~al.}(2019)Mantz, Allen, Battaglia, Benson, Canning, Ettori,
  Evrard, {von der Linden}, McDonald, Abidi, Ahmed, Amin, Ansarinejad,
  Armstrong, Avestruz, Baccigalupi, Bandura, Barkhouse, moni Basu, Bavdhankar,
  Bender, {de Bernardis}, Bischoff, Biviano, Bleem, Bocquet, Bond, Borgani,
  Borrill, Boutigny, Frye, Bruggen, Cai, Carlstrom, Castander, Challinor,
  Clowe, Cohn, Comparat, Cooray, Coulton, {Cyr-Racine}, Daddi, Delabrouille,
  Dell'antonio, Demarteau, Donahue, Dunkley, Escoffier, {Essinger-Hileman},
  Fabbian, Fabjan, Farahi, Foreman, Fraisse, Garcia, Gaspari, Gerbino, Gitti,
  Gluscevic, Gonzalez, G{\'o}rski, Gruen, Gudmundsson, Gupta, {de Haan},
  Hernquist, Hirata, Hlozek, Jeltema, {Cohen-Tanugi}, Johnson, Kadota,
  Kamionkowski, Khatri, Kisner, Kneib, Knox, Kovetz, Krause, Lattanzi, Lau,
  Liguori, Lovisari, de~la Macorra, Masi, Masui, Maughan, Maurogordato,
  McMahon, McNamara, Melchior, Mertens, Meyers, Mirbabayi, More, Motloch,
  Moustakas, Mroczkowski, Mukherjee, Nagai, Nagy, Naselsky, Nati, Newburgh,
  Niemack, Nomerotski, Noordeh, Ntampaka, Ota, Page, Palmese, {Penna-Lima},
  Piacentni, Pierpaoli, Plazas, Pogosian, Pointecouteau, Prakash, Pratt,
  {Prescod-Weinstein}, Pryke, Puglisi, Rapetti, Raveri, Reichardt, Reiprich,
  Remazeilles, Rhodes, Ricci, Rocha, Rose, Rozo, Ruhl, Sadun, Saliwanchik,
  Schaan, Schmidt, Fromenteau, Sehgal, Senatore, Seo, Sereno, Shafieloo, Shan,
  Shandera, Sherwin, Simon, Sridhar, Staggs, Stern, Suzuki, Tsai, Turriziani,
  Umilta, Vazza, Vieregg, Vikhlinin, Walker, Watson, {van Weeren}, Weller,
  Werner, Whitehorn, Wong, Wright, Wu, Xu, Yasini, Zemcov, Zhang, Zhao, Zheng,
  Zhu, Zhuravleva, Zuntz, Hickox, Churazov, Nulsen, Jones, Wang, \&
  Desai}]{mantzFutureLandscapeHighRedshift2019}
Mantz, A., Allen, S.~W., Battaglia, N., {et~al.} 2019, Bulletin of the American
  Astronomical Society, 51, 279, \dodoi{10.48550/arXiv.1903.05606}

\bibitem[{Mao \& Schneider(1998)}]{maoEvidenceSubstructureLens1998}
Mao, S., \& Schneider, P. 1998, Monthly Notices of the Royal Astronomical
  Society, 295, 587, \dodoi{10.1046/j.1365-8711.1998.01319.x}

\bibitem[{Meneghetti(2021)}]{meneghettiIntroductionGravitationalLensing2021}
Meneghetti, M. 2021, Introduction to {{Gravitational Lensing}}: {{With Python
  Examples}}, 1st edn. (Cham, Switzerland: Springer)

\bibitem[{Meneghetti {et~al.}(2017)Meneghetti, Natarajan, Coe, Contini,
  De~Lucia, Giocoli, Acebron, Borgani, Bradac, Diego, Hoag, Ishigaki, Johnson,
  Jullo, Kawamata, Lam, Limousin, Liesenborgs, Oguri, Sebesta, Sharon,
  Williams, \& Zitrin}]{meneghettiFrontierFieldsLens2017}
Meneghetti, M., Natarajan, P., Coe, D., {et~al.} 2017, Monthly Notices of the
  Royal Astronomical Society, 472, 3177, \dodoi{10.1093/mnras/stx2064}

\bibitem[{Meneghetti {et~al.}(2020)Meneghetti, Davoli, Bergamini, Rosati,
  Natarajan, Giocoli, Caminha, Metcalf, Rasia, Borgani, Calura, Grillo,
  Mercurio, \& Vanzella}]{meneghettiExcessSmallscaleGravitational2020}
Meneghetti, M., Davoli, G., Bergamini, P., {et~al.} 2020, Science, 369, 1347,
  \dodoi{10.1126/science.aax5164}

\bibitem[{Meneghetti {et~al.}(2022)Meneghetti, Ragagnin, Borgani, Calura,
  Despali, Giocoli, Granato, Grillo, Moscardini, Rasia, Rosati, Angora,
  Bassini, Bergamini, Caminha, Granata, Mercurio, Metcalf, Natarajan, Nonino,
  Pignataro, {Ragone-Figueroa}, Vanzella, Acebron, Dolag, Murante, Taffoni,
  Tornatore, Tortorelli, \&
  Valentini}]{meneghettiProbabilityGalaxygalaxyStrong2022}
Meneghetti, M., Ragagnin, A., Borgani, S., {et~al.} 2022, The Probability of
  Galaxy-Galaxy Strong Lensing Events in Hydrodynamical Simulations of Galaxy
  Clusters,  arXiv.
\newblock \doeprint{2204.09065}

\bibitem[{Meneghetti {et~al.}(2023)Meneghetti, Cui, Rasia, Yepes, Acebron,
  Angora, Bergamini, Borgani, Calura, Despali, Giocoli, Granata, Grillo, Knebe,
  Macci{\`o}, Mercurio, Moscardini, Natarajan, Ragagnin, Rosati, \&
  Vanzella}]{meneghettiPersistentExcessGalaxygalaxy2023}
Meneghetti, M., Cui, W., Rasia, E., {et~al.} 2023, Astronomy and Astrophysics,
  678, L2, \dodoi{10.1051/0004-6361/202346975}

\bibitem[{Mercier {et~al.}(2023)Mercier, Shuntov, Gavazzi, Nightingale, Arango,
  Ilbert, Amvrosiadis, Ciesla, Casey, Jin, Faisst, Andika, Drakos, Enia,
  Franco, Gillman, Gozaliasl, Hayward, {Huertas-Company}, Kartaltepe,
  Koekemoer, Laigle, Le~Borgne, Magdis, Mahler, Maraston, Martin, Massey,
  McCracken, Moutard, Paquereau, Rhodes, Robertson, Sanders, Trebitsch, Tresse,
  \& Vijayan}]{mercierCOSMOSWebRingIndepth2023}
Mercier, W., Shuntov, M., Gavazzi, R., {et~al.} 2023, The {{COSMOS-Web}} Ring:
  In-Depth Characterization of an {{Einstein}} Ring Lensing System at
  Z{\textasciitilde}2, \dodoi{10.48550/arXiv.2309.15986}

\bibitem[{Merten {et~al.}(2011)Merten, Coe, Dupke, Massey, Zitrin, Cypriano,
  Okabe, Frye, Braglia, {Jim{\'e}nez-Teja}, Ben{\'i}tez, Broadhurst, Rhodes,
  Meneghetti, Moustakas, Sodr{\'e}~Jr, Krick, \&
  Bregman}]{mertenCreationCosmicStructure2011}
Merten, J., Coe, D., Dupke, R., {et~al.} 2011, Monthly Notices of the Royal
  Astronomical Society, 417, 333, \dodoi{10.1111/j.1365-2966.2011.19266.x}

\bibitem[{Minor {et~al.}(2021)Minor, {Gad-Nasr}, Kaplinghat, \&
  Vegetti}]{minorUnexpectedHighConcentration2021}
Minor, Q., {Gad-Nasr}, S., Kaplinghat, M., \& Vegetti, S. 2021, Monthly Notices
  of the Royal Astronomical Society, 507, 1662, \dodoi{10.1093/mnras/stab2247}

\bibitem[{Minor {et~al.}(2017)Minor, Kaplinghat, \&
  Li}]{minorRobustMassEstimator2017}
Minor, Q.~E., Kaplinghat, M., \& Li, N. 2017, ApJ, 845, 118,
  \dodoi{10.3847/1538-4357/aa7fee}

\bibitem[{Moore {et~al.}(1999)Moore, Ghigna, Governato, Lake, Quinn, Stadel, \&
  Tozzi}]{mooreDarkMatterSubstructure1999}
Moore, B., Ghigna, S., Governato, F., {et~al.} 1999, The Astrophysical Journal,
  524, L19, \dodoi{10.1086/312287}

\bibitem[{Mu{\~n}oz {et~al.}(2001)Mu{\~n}oz, Kochanek, \&
  Keeton}]{munozCuspedMassModels2001}
Mu{\~n}oz, J.~A., Kochanek, C.~S., \& Keeton, C.~R. 2001, ApJ, 558, 657,
  \dodoi{10.1086/322314}

\bibitem[{Nadler {et~al.}(2023)Nadler, Yang, \&
  Yu}]{nadlerSelfinteractingDarkMatter2023}
Nadler, E.~O., Yang, D., \& Yu, H.-B. 2023, ApJL, 958, L39,
  \dodoi{10.3847/2041-8213/ad0e09}

\bibitem[{Natarajan \& Kneib(1997)}]{natarajanLensingGalaxyHaloes1997}
Natarajan, P., \& Kneib, J.-P. 1997, Monthly Notices of the Royal Astronomical
  Society, 287, 833, \dodoi{10.1093/mnras/287.4.833}

\bibitem[{Natarajan {et~al.}(2002)Natarajan, Kneib, \&
  Smail}]{natarajanEvidenceTidalStripping2002}
Natarajan, P., Kneib, J.-P., \& Smail, I. 2002, The Astrophysical Journal, 580,
  L11, \dodoi{10.1086/345399}

\bibitem[{Natarajan {et~al.}(2009)Natarajan, Kneib, Smail, Treu, Ellis, Moran,
  Limousin, \& Czoske}]{natarajanSurvivalDarkMatter2009}
Natarajan, P., Kneib, J.-P., Smail, I., {et~al.} 2009, The Astrophysical
  Journal, 693, 970, \dodoi{10.1088/0004-637X/693/1/970}

\bibitem[{Natarajan {et~al.}(2017)Natarajan, Chadayammuri, Jauzac, Richard,
  Kneib, Ebeling, Jiang, {van den Bosch}, Limousin, Jullo, Atek, Pillepich,
  Popa, Marinacci, Hernquist, Meneghetti, \&
  Vogelsberger}]{natarajanMappingSubstructureHST2017}
Natarajan, P., Chadayammuri, U., Jauzac, M., {et~al.} 2017, Monthly Notices of
  the Royal Astronomical Society, 468, 1962, \dodoi{10.1093/mnras/stw3385}

\bibitem[{Navarro {et~al.}(1997)Navarro, Frenk, \&
  White}]{navarroUniversalDensityProfile1997}
Navarro, J.~F., Frenk, C.~S., \& White, S. D.~M. 1997, The Astrophysical
  Journal, 490, 493, \dodoi{10.1086/304888}

\bibitem[{Oguri {et~al.}(2009)Oguri, Hennawi, Gladders, Dahle, Natarajan,
  Dalal, Koester, Sharon, \& Bayliss}]{oguriSubaruWeakLensing2009}
Oguri, M., Hennawi, J.~F., Gladders, M.~D., {et~al.} 2009, The Astrophysical
  Journal, 699, 1038, \dodoi{10.1088/0004-637X/699/2/1038}

\bibitem[{Ostdiek {et~al.}(2022)Ostdiek, Rivero, \&
  Dvorkin}]{ostdiekExtractingSubhaloMass2022}
Ostdiek, B., Rivero, A.~D., \& Dvorkin, C. 2022, ApJ, 927, 83,
  \dodoi{10.3847/1538-4357/ac2d8d}

\bibitem[{Perivolaropoulos \&
  Skara(2022)}]{perivolaropoulosChallengesLCDMUpdate2022}
Perivolaropoulos, L., \& Skara, F. 2022, New Astronomy Reviews, 95, 101659,
  \dodoi{10.1016/j.newar.2022.101659}

\bibitem[{Petts {et~al.}(2015)Petts, Gualandris, \&
  Read}]{pettsSemianalyticDynamicalFriction2015}
Petts, J.~A., Gualandris, A., \& Read, J.~I. 2015, Mon. Not. R. Astron. Soc.,
  454, 3778, \dodoi{10.1093/mnras/stv2235}

\bibitem[{Pignataro {et~al.}(2021)Pignataro, Bergamini, Meneghetti, Vanzella,
  Calura, Grillo, Rosati, Angora, Brammer, Caminha, Mercurio, Nonino, \&
  Tozzi}]{pignataroStrongLensingModel2021}
Pignataro, G.~V., Bergamini, P., Meneghetti, M., {et~al.} 2021, A\&A, 655, A81,
  \dodoi{10.1051/0004-6361/202141586}

\bibitem[{{Planck Collaboration} {et~al.}(2014){Planck Collaboration}, Ade,
  Aghanim, {Armitage-Caplan}, Arnaud, Ashdown, {Atrio-Barandela}, Aumont,
  Aussel, Baccigalupi, Banday, Barreiro, Barrena, Bartelmann, Bartlett,
  Battaner, Benabed, Beno{\^i}t, {Benoit-L{\'e}vy}, Bernard, Bersanelli,
  Bielewicz, Bikmaev, Bobin, Bock, B{\"o}hringer, Bonaldi, Bond, Borrill,
  Bouchet, Bridges, Bucher, Burenin, Burigana, Butler, Cardoso, Carvalho,
  Catalano, Challinor, Chamballu, Chary, Chen, Chiang, Chiang, Chon,
  Christensen, Churazov, Church, Clements, Colombi, Colombo, Comis, Couchot,
  Coulais, Crill, Curto, Cuttaia, Da~Silva, Dahle, Danese, Davies, Davis, {de
  Bernardis}, {de Rosa}, {de Zotti}, Delabrouille, Delouis, D{\'e}mocl{\`e}s,
  D{\'e}sert, Dickinson, Diego, Dolag, Dole, Donzelli, Dor{\'e}, Douspis,
  Dupac, Efstathiou, Eisenhardt, En{\ss}lin, Eriksen, Feroz, Finelli,
  {Flores-Cacho}, Forni, Frailis, Franceschi, Fromenteau, Galeotta, Ganga,
  {G{\'e}nova-Santos}, Giard, Giardino, Gilfanov, {Giraud-H{\'e}raud},
  {Gonz{\'a}lez-Nuevo}, G{\'o}rski, Grainge, Gratton, Gregorio, Groeneboom,
  Gruppuso, Hansen, Hanson, Harrison, Hempel, {Henrot-Versill{\'e}},
  {Hern{\'a}ndez-Monteagudo}, Herranz, Hildebrandt, Hivon, Hobson, Holmes,
  Hornstrup, Hovest, Huffenberger, Hurier, {Hurley-Walker}, Jaffe, Jaffe,
  Jones, Juvela, Keih{\"a}nen, Keskitalo, Khamitov, Kisner, Kneissl, Knoche,
  Knox, Kunz, {Kurki-Suonio}, Lagache, L{\"a}hteenm{\"a}ki, Lamarre, Lasenby,
  Laureijs, Lawrence, Leahy, Leonardi, {Le{\'o}n-Tavares}, Lesgourgues, Li,
  Liddle, Liguori, Lilje, {Linden-V{\o}rnle}, {L{\'o}pez-Caniego}, Lubin,
  {Mac{\'i}as-P{\'e}rez}, MacTavish, Maffei, Maino, Mandolesi, Maris, Marshall,
  Martin, {Mart{\'i}nez-Gonz{\'a}lez}, Masi, Massardi, Matarrese, Matthai,
  Mazzotta, Mei, Meinhold, Melchiorri, Melin, Mendes, Mennella, Migliaccio,
  Mikkelsen, Mitra, {Miville-Desch{\^e}nes}, Moneti, Montier, Morgante,
  Mortlock, Munshi, Murphy, Naselsky, Nati, Natoli, Nesvadba, Netterfield,
  {N{\o}rgaard-Nielsen}, Noviello, Novikov, Novikov, O'Dwyer, Olamaie, Osborne,
  Oxborrow, Paci, Pagano, Pajot, Paoletti, Pasian, Patanchon, Pearson,
  Perdereau, Perotto, Perrott, Perrotta, Piacentini, Piat, Pierpaoli,
  Pietrobon, Plaszczynski, Pointecouteau, Polenta, Ponthieu, Popa, Poutanen,
  Pratt, Pr{\'e}zeau, Prunet, Puget, Rachen, Reach, Rebolo, Reinecke,
  Remazeilles, Renault, Ricciardi, Riller, Ristorcelli, Rocha, Rosset, Roudier,
  {Rowan-Robinson}, {Rubi{\~n}o-Mart{\'i}n}, Rumsey, Rusholme, Sandri, Santos,
  Saunders, Savini, Schammel, Scott, Seiffert, Shellard, Shimwell, Spencer,
  Stanford, Starck, Stolyarov, Stompor, Sudiwala, Sunyaev, Sureau, Sutton,
  {Suur-Uski}, Sygnet, Tauber, Tavagnacco, Terenzi, Toffolatti, Tomasi,
  Tristram, Tucci, Tuovinen, T{\"u}rler, Umana, Valenziano, Valiviita,
  Van~Tent, Vibert, Vielva, Villa, Vittorio, Wade, Wandelt, White, White, Yvon,
  Zacchei, \& Zonca}]{planckcollaborationPlanck2013Results2014}
{Planck Collaboration}, Ade, P. A.~R., Aghanim, N., {et~al.} 2014, A\&A, 571,
  A29, \dodoi{10.1051/0004-6361/201321523}

\bibitem[{Pontzen \& Governato(2012)}]{pontzenHowSupernovaFeedback2012}
Pontzen, A., \& Governato, F. 2012, Monthly Notices of the Royal Astronomical
  Society, 421, 3464, \dodoi{10.1111/j.1365-2966.2012.20571.x}

\bibitem[{Postman {et~al.}(2012)Postman, Coe, Ben{\'i}tez, Bradley, Broadhurst,
  Donahue, Ford, Graur, Graves, Jouvel, Koekemoer, Lemze, Medezinski, Molino,
  Moustakas, Ogaz, Riess, Rodney, Rosati, Umetsu, Zheng, Zitrin, Bartelmann,
  Bouwens, Czakon, Golwala, Host, Infante, Jha, {Jimenez-Teja}, Kelson, Lahav,
  Lazkoz, Maoz, McCully, Melchior, Meneghetti, Merten, Moustakas, Nonino,
  Patel, Reg{\"o}s, Sayers, Seitz, \& {Van der
  Wel}}]{postmanCLUSTERLENSINGSUPERNOVA2012}
Postman, M., Coe, D., Ben{\'i}tez, N., {et~al.} 2012, ApJS, 199, 25,
  \dodoi{10.1088/0067-0049/199/2/25}

\bibitem[{Ragagnin {et~al.}(2022)Ragagnin, Meneghetti, Bassini,
  {Ragone-Figueroa}, Luigi~Granato, Despali, Giocoli, Granata, Moscardini,
  Bergamini, Rasia, Valentini, Borgani, Calura, Dolag, Grillo, Mercurio,
  Murante, Natarajan, Rosati, Taffoni, Tornatore, \&
  Tortorelli}]{ragagninGalaxiesCentralRegions2022}
Ragagnin, A., Meneghetti, M., Bassini, L., {et~al.} 2022, A\&A, 665, A16,
  \dodoi{10.1051/0004-6361/202243651}

\bibitem[{{Ragone-Figueroa} {et~al.}(2018){Ragone-Figueroa}, Granato, Ferraro,
  Murante, Biffi, Borgani, Planelles, \&
  Rasia}]{ragone-figueroaBCGMassEvolution2018}
{Ragone-Figueroa}, C., Granato, G.~L., Ferraro, M.~E., {et~al.} 2018, Monthly
  Notices of the Royal Astronomical Society, 479, 1125,
  \dodoi{10.1093/mnras/sty1639}

\bibitem[{Rahaman {et~al.}(2021)Rahaman, Raja, Datta, Burns, Alden, \&
  Rapetti}]{rahamanXrayRadioStudy2021}
Rahaman, M., Raja, R., Datta, A., {et~al.} 2021, Monthly Notices of the Royal
  Astronomical Society, 505, 480, \dodoi{10.1093/mnras/stab1225}

\bibitem[{Richard {et~al.}(2021)Richard, Claeyssens, Lagattuta, Guaita, Bauer,
  Pello, Carton, Bacon, Soucail, Lyon, Kneib, Mahler, Cl{\'e}ment, Mercier,
  Variu, Tamone, Ebeling, Schmidt, Nanayakkara, Maseda, Weilbacher, Bouch{\'e},
  Bouwens, Wisotzki, {de la Vieuville}, Martinez, \&
  Patr{\'i}cio}]{richardAtlasMUSEObservations2021}
Richard, J., Claeyssens, A., Lagattuta, D., {et~al.} 2021, Astronomy and
  Astrophysics, 646, A83, \dodoi{10.1051/0004-6361/202039462}

\bibitem[{{Rivera-Thorsen} {et~al.}(2017){Rivera-Thorsen}, Dahle, Gronke,
  Bayliss, Rigby, Simcoe, Bordoloi, Turner, \&
  Furesz}]{rivera-thorsenSunburstArcDirect2017}
{Rivera-Thorsen}, T.~E., Dahle, H., Gronke, M., {et~al.} 2017, Astronomy and
  Astrophysics, 608, L4, \dodoi{10.1051/0004-6361/201732173}

\bibitem[{Robertson(2021)}]{robertsonGalaxyGalaxyStrong2021}
Robertson, A. 2021, Monthly Notices of the Royal Astronomical Society: Letters,
  504, L7, \dodoi{10.1093/mnrasl/slab028}

\bibitem[{Rosati {et~al.}(2014)Rosati, Balestra, Grillo, Mercurio, Nonino,
  Biviano, Girardi, Vanzella, \& {Clash-VLT
  Team}}]{rosatiCLASHVLTVIMOSLarge2014}
Rosati, P., Balestra, I., Grillo, C., {et~al.} 2014, The Messenger, 158, 48

\bibitem[{Sand {et~al.}(2004)Sand, Treu, Smith, \&
  Ellis}]{sandDarkMatterDistribution2004}
Sand, D.~J., Treu, T., Smith, G.~P., \& Ellis, R.~S. 2004, The Astrophysical
  Journal, 604, 88, \dodoi{10.1086/382146}

\bibitem[{Sartoris {et~al.}(2016)Sartoris, Biviano, Fedeli, Bartlett, Borgani,
  Costanzi, Giocoli, Moscardini, Weller, Ascaso, Bardelli, Maurogordato, \&
  Viana}]{sartorisNextGenerationCosmology2016}
Sartoris, B., Biviano, A., Fedeli, C., {et~al.} 2016, Monthly Notices of the
  Royal Astronomical Society, 459, 1764, \dodoi{10.1093/mnras/stw630}

\bibitem[{Shan {et~al.}(2010)Shan, Qin, Fort, Tao, Wu, \&
  Zhao}]{shanOffsetDarkMatter2010}
Shan, H., Qin, B., Fort, B., {et~al.} 2010, Monthly Notices of the Royal
  Astronomical Society, 406, 1134, \dodoi{10.1111/j.1365-2966.2010.16739.x}

\bibitem[{Sharon {et~al.}(2022)Sharon, Mahler, {Rivera-Thorsen}, Dahle,
  Gladders, Bayliss, Florian, Kim, Khullar, Mainali, Napier, Navarre, Rigby,
  Gonz{\'a}lez, \& Sharma}]{sharonCosmicTelescopeThat2022}
Sharon, K., Mahler, G., {Rivera-Thorsen}, T.~E., {et~al.} 2022, ApJ, 941, 203,
  \dodoi{10.3847/1538-4357/ac927a}

\bibitem[{Spergel {et~al.}(2013)Spergel, Gehrels, Breckinridge, Donahue,
  Dressler, Gaudi, Greene, Guyon, Hirata, Kalirai, Kasdin, Moos, Perlmutter,
  Postman, Rauscher, Rhodes, Wang, Weinberg, Centrella, Traub, Baltay, Colbert,
  Bennett, Kiessling, Macintosh, Merten, Mortonson, Penny, Rozo, Savransky,
  Stapelfeldt, Zu, Baker, Cheng, Content, Dooley, Foote, Goullioud, Grady,
  Jackson, Kruk, Levine, Melton, Peddie, Ruffa, \&
  Shaklan}]{spergelWFIRST2WhatEvery2013}
Spergel, D., Gehrels, N., Breckinridge, J., {et~al.} 2013, {{WFIRST-2}}.4:
  {{What Every Astronomer Should Know}},  arXiv,
  \dodoi{10.48550/arXiv.1305.5425}

\bibitem[{Spergel \&
  Steinhardt(2000)}]{spergelObservationalEvidenceSelfInteracting2000}
Spergel, D.~N., \& Steinhardt, P.~J. 2000, Phys. Rev. Lett., 84, 3760,
  \dodoi{10.1103/PhysRevLett.84.3760}

\bibitem[{Springel(2005)}]{springelCosmologicalSimulationCode2005}
Springel, V. 2005, Monthly Notices of the Royal Astronomical Society, 364,
  1105, \dodoi{10.1111/j.1365-2966.2005.09655.x}

\bibitem[{Srivastava {et~al.}(2024)Srivastava, Cui, Meneghetti, Dave, Knebe,
  Ragagnin, Giocoli, Calura, Despali, Moscardini, \&
  Yepes}]{srivastavaThreeHundredMsub2024}
Srivastava, A., Cui, W., Meneghetti, M., {et~al.} 2024, Monthly Notices of the
  Royal Astronomical Society, 528, 4451, \dodoi{10.1093/mnras/stae320}

\bibitem[{Steinhardt {et~al.}(2020)Steinhardt, Jauzac, Acebron, Atek, Capak,
  Davidzon, Eckert, Harvey, Koekemoer, Lagos, Mahler, Montes, Niemiec, Nonino,
  Oesch, Richard, Rodney, Schaller, Sharon, Strolger, Allingham, Amara,
  Bah{\'e}, B{\oe}hm, Bose, Bouwens, Bradley, Brammer, Broadhurst, Ca{\~n}as,
  Cen, Cl{\'e}ment, Clowe, Coe, Connor, Darvish, Diego, Ebeling, Edge, Egami,
  Ettori, Faisst, Frye, Furtak, {G{\'o}mez-Guijarro}, Gonz{\'a}lez, Gonzalez,
  Graur, Gruen, Harvey, Hensley, {Hovis-Afflerbach}, Jablonka, Jha, Jullo,
  Kneib, Kokorev, Lagattuta, Limousin, von~der Linden, Linzer, Lopez, Magdis,
  Massey, Masters, Maturi, McCully, McGee, Meneghetti, Mobasher, Moustakas,
  Murphy, Natarajan, Neyrinck, O'Connor, Oguri, Pagul, Rhodes, Rich, Robertson,
  Sereno, Shan, Smith, Sneppen, Squires, Tam, Tchernin, Toft, Umetsu, Weaver,
  van Weeren, Williams, Wilson, Yan, \&
  Zitrin}]{steinhardtBUFFALOHSTSurvey2020}
Steinhardt, C.~L., Jauzac, M., Acebron, A., {et~al.} 2020, ApJS, 247, 64,
  \dodoi{10.3847/1538-4365/ab75ed}

\bibitem[{Umetsu {et~al.}(2016)Umetsu, Zitrin, Gruen, Merten, Donahue, \&
  Postman}]{umetsuCLASHJOINTANALYSIS2016}
Umetsu, K., Zitrin, A., Gruen, D., {et~al.} 2016, ApJ, 821, 116,
  \dodoi{10.3847/0004-637X/821/2/116}

\bibitem[{Umetsu {et~al.}(2012)Umetsu, Medezinski, Nonino, Merten, Zitrin,
  Molino, Grillo, Carrasco, Donahue, Mahdavi, Coe, Postman, Koekemoer, Czakon,
  Sayers, Mroczkowski, Golwala, Koch, Lin, Molnar, Rosati, Balestra, Mercurio,
  Scodeggio, Biviano, Anguita, Infante, Seidel, Sendra, Jouvel, Host, Lemze,
  Broadhurst, Meneghetti, Moustakas, Bartelmann, Ben{\'i}tez, Bouwens, Bradley,
  Ford, {Jim{\'e}nez-Teja}, Kelson, Lahav, Melchior, Moustakas, Ogaz, Seitz, \&
  Zheng}]{umetsuCLASHMassDistribution2012}
Umetsu, K., Medezinski, E., Nonino, M., {et~al.} 2012, The Astrophysical
  Journal, 755, 56, \dodoi{10.1088/0004-637X/755/1/56}

\bibitem[{{van Dokkum} {et~al.}(2023){van Dokkum}, Brammer, Wang, Leja, \&
  Conroy}]{vandokkumMassiveCompactQuiescent2023}
{van Dokkum}, P., Brammer, G., Wang, B., Leja, J., \& Conroy, C. 2023, Nat
  Astron, 1, \dodoi{10.1038/s41550-023-02103-9}

\bibitem[{Vegetti {et~al.}(2014)Vegetti, Koopmans, Auger, Treu, \&
  Bolton}]{vegettiInferenceColdDark2014}
Vegetti, S., Koopmans, L. V.~E., Auger, M.~W., Treu, T., \& Bolton, A.~S. 2014,
  Monthly Notices of the Royal Astronomical Society, 442, 2017,
  \dodoi{10.1093/mnras/stu943}

\bibitem[{Vegetti \& Vogelsberger(2014)}]{vegettiDensityProfileDark2014}
Vegetti, S., \& Vogelsberger, M. 2014, Monthly Notices of the Royal
  Astronomical Society, 442, 3598, \dodoi{10.1093/mnras/stu1284}

\bibitem[{Vegetti {et~al.}(2023)Vegetti, Birrer, Despali, Fassnacht, Gilman,
  Hezaveh, Perreault~Levasseur, McKean, Powell, O'Riordan, \&
  Vernardos}]{vegettiStrongGravitationalLensing2023}
Vegetti, S., Birrer, S., Despali, G., {et~al.} 2023, Strong Gravitational
  Lensing as a Probe of Dark Matter, \dodoi{10.48550/arXiv.2306.11781}

\bibitem[{{Wagner-Carena} {et~al.}(2023){Wagner-Carena}, Aalbers, Birrer,
  Nadler, {Darragh-Ford}, Marshall, \&
  Wechsler}]{wagner-carenaImagesDarkMatter2023}
{Wagner-Carena}, S., Aalbers, J., Birrer, S., {et~al.} 2023, ApJ, 942, 75,
  \dodoi{10.3847/1538-4357/aca525}

\bibitem[{Wechsler {et~al.}(2002)Wechsler, Bullock, Primack, Kravtsov, \&
  Dekel}]{wechslerConcentrationsDarkHalos2002}
Wechsler, R.~H., Bullock, J.~S., Primack, J.~R., Kravtsov, A.~V., \& Dekel, A.
  2002, The Astrophysical Journal, 568, 52, \dodoi{10.1086/338765}

\bibitem[{Weinberg {et~al.}(2015)Weinberg, Bullock, Governato, {Kuzio de
  Naray}, \& Peter}]{weinbergColdDarkMatter2015}
Weinberg, D.~H., Bullock, J.~S., Governato, F., {Kuzio de Naray}, R., \& Peter,
  A. H.~G. 2015, Proceedings of the National Academy of Science, 112, 12249,
  \dodoi{10.1073/pnas.1308716112}

\bibitem[{Williamson {et~al.}(2011)Williamson, Benson, High, Vanderlinde, Ade,
  Aird, Andersson, Armstrong, Ashby, Bautz, Bazin, Bertin, Bleem, Bonamente,
  Brodwin, Carlstrom, Chang, Chapman, Clocchiatti, Crawford, Crites, {de Haan},
  Desai, Dobbs, Dudley, Fazio, Foley, Forman, Garmire, George, Gladders,
  Gonzalez, Halverson, Holder, Holzapfel, Hoover, Hrubes, Jones, Joy, Keisler,
  Knox, Lee, Leitch, Lueker, {Luong-Van}, Marrone, McMahon, Mehl, Meyer, Mohr,
  Montroy, Murray, Padin, Plagge, Pryke, Reichardt, Rest, Ruel, Ruhl,
  Saliwanchik, Saro, Schaffer, Shaw, Shirokoff, Song, Spieler, Stalder,
  Stanford, Staniszewski, Stark, Story, Stubbs, Vieira, Vikhlinin, \&
  Zenteno}]{williamsonSUNYAEVZELDOVICHSELECTEDSAMPLE2011}
Williamson, R., Benson, B.~A., High, F.~W., {et~al.} 2011, ApJ, 738, 139,
  \dodoi{10.1088/0004-637X/738/2/139}

\bibitem[{Yang \& Yu(2021)}]{yangSelfinteractingDarkMatter2021}
Yang, D., \& Yu, H.-B. 2021, Phys. Rev. D, 104, 103031,
  \dodoi{10.1103/PhysRevD.104.103031}

\bibitem[{Zaritsky \& Behroozi(2023)}]{zaritskyPhotometricMassEstimation2023}
Zaritsky, D., \& Behroozi, P. 2023, Monthly Notices of the Royal Astronomical
  Society, 519, 871, \dodoi{10.1093/mnras/stac3610}

\bibitem[{Zeng {et~al.}(2022)Zeng, Peter, Du, Benson, Kim, Jiang, {Cyr-Racine},
  \& Vogelsberger}]{zengCorecollapseEvaporationTidal2022}
Zeng, Z.~C., Peter, A. H.~G., Du, X., {et~al.} 2022, Monthly Notices of the
  Royal Astronomical Society, 513, 4845, \dodoi{10.1093/mnras/stac1094}

\bibitem[{Zhao {et~al.}(2009)Zhao, Jing, Mo, \&
  B{\"o}rner}]{zhaoAccurateUniversalModels2009}
Zhao, D.~H., Jing, Y.~P., Mo, H.~J., \& B{\"o}rner, G. 2009, The Astrophysical
  Journal, 707, 354, \dodoi{10.1088/0004-637X/707/1/354}

\bibitem[{Zhao {et~al.}(2003)Zhao, Mo, Jing, \&
  B{\"o}rner}]{zhaoGrowthStructureDark2003}
Zhao, D.~H., Mo, H.~J., Jing, Y.~P., \& B{\"o}rner, G. 2003, Monthly Notices of
  the Royal Astronomical Society, 339, 12,
  \dodoi{10.1046/j.1365-8711.2003.06135.x}

\end{thebibliography}
